\documentclass[prb,twocolumn,superscriptaddress]{revtex4-1}  

\usepackage{tikz}
\newcommand*\circled[1]{\tikz[baseline=(char.base)]{
            \node[shape=circle,draw,inner sep=1pt,fill=black] (char) {#1};}}

\usepackage[sort&compress]{natbib}
\usepackage{soul}
\usepackage{graphicx}
\usepackage{verbatim}
\usepackage{float}
\usepackage{sidecap}
\sidecaptionvpos{figure}{c}
\usepackage{lipsum}
\usepackage{setspace}
\usepackage{dcolumn}
\usepackage{bm}
\usepackage{amssymb}
\usepackage{amsmath}	
\usepackage{color}
\usepackage{multirow}
\usepackage{lipsum}
\usepackage[a4paper]{geometry}
\newgeometry{top=1.5cm,right=1.25cm,left=1.25cm,bottom=1.5cm}
\usepackage[colorlinks=true, urlcolor=blue, pdfborder={0 0 0}]{hyperref}
\hypersetup{
linkcolor=blue,
citecolor=blue
}

\begin{document}

\title{\large{Spatial Metrology of Dopants in Silicon with Exact Lattice Site Precision}}

\author{M. Usman} \email{musman@unimelb.edu.au} \affiliation{Centre for Quantum Computation and Communication Technology, School of Physics, The University of Melbourne, Parkville, 3010, VIC, Australia.}
\author{J. Bocquel} \affiliation{Centre for Quantum Computation and Communication Technology, School of Physics, The University of New South Wales, Sydney, 2052, NSW, Australia.} 
\author{J. Salfi} \affiliation{Centre for Quantum Computation and Communication Technology, School of Physics, The University of New South Wales, Sydney, 2052, NSW, Australia.} 
\author{B. Voisin} \affiliation{Centre for Quantum Computation and Communication Technology, School of Physics, The University of New South Wales, Sydney, 2052, NSW, Australia.} 
\author{A. Tankasala}  \affiliation{Electrical and Computer Engineering Department, Purdue University, West Lafayette, Indiana, USA.}
\author{R. Rahman}  \affiliation{Electrical and Computer Engineering Department, Purdue University, West Lafayette, Indiana, USA.}
\author{M. Y. Simmons}  \affiliation{Centre for Quantum Computation and Communication Technology, School of Physics, The University of New South Wales, Sydney, 2052, NSW, Australia.} 
\author{S. Rogge}  \affiliation{Centre for Quantum Computation and Communication Technology, School of Physics, The University of New South Wales, Sydney, 2052, NSW, Australia.}  
\author{L. C. L. Hollenberg} \email{lloydch@unimelb.edu.au} \affiliation{Centre for Quantum Computation and Communication Technology, School of Physics, The University of Melbourne, Parkville, 3010, VIC, Australia.}

\begin{abstract}
\noindent
\textbf{\small{The aggressive scaling of silicon-based nanoelectronics has reached the regime where device function is affected not only by the presence of individual dopants, but more
critically their position in the structure. The quantitative determination of the positions of subsurface dopant atoms is an important issue in a range of applications from channel doping in ultra-scaled transistors to quantum information processing, and hence poses a significant challenge. Here, we establish a metrology combining low-temperature scanning tunnelling microscopy (STM) imaging and a comprehensive quantum treatment of the dopant-STM system to pin-point the exact lattice-site location of sub-surface dopants in silicon. The technique is underpinned by the observation that STM images of sub-surface dopants typically contain many atomic-sized features in ordered patterns, which are highly sensitive to the details of the STM tip orbital and the absolute lattice-site position of the dopant atom itself. We demonstrate the technique on two types of dopant samples in silicon -- the first where phosphorus dopants are placed with high precision, and a second containing randomly placed arsenic dopants. Based on the quantitative agreement between STM measurements and multi-million-atom calculations, the precise lattice site of these dopants is determined, demonstrating that the metrology works to depths of about 36 lattice planes. The ability to uniquely determine the exact positions of sub–surface dopants down to depths of 5 nm will provide critical knowledge in the design and optimisation of nanoscale devices for both classical and quantum computing applications.}}
\end{abstract}
\maketitle

As we approach the ultimate regime of Feynman's vision~\footnote{Feynman's fancy, Chemistry World 58, January 2009, Royal Society of Chemistry (RSC)} of nanotechnology based on atom-by-atom fabrication~\cite{Fuechsle_NN_2012, Weber_Science_2012, Ho_NM_2008, SAE_2013}, there is a critical need to match advances in miniaturisation with atomically precise metrology. In conventional CMOS~\cite{Pierre_NNano_2010} and tunnelling field effect~\cite{Sarkar_Nature_2015, Ionescu_Nature_2011} transistors the key relationship between doping profile and performance is now dominated by the positions of just a few dopant atoms, and currently cannot be quantitatively determined. Beyond conventional nanoelectronic devices, in quantum processors based on phosphorus dopants in silicon~\cite{Hill_science_2015} the precise locations of the individual dopants is critical to the design and operation of spin-based quantum logic gates. Previous studies on locating subsurface dopant positions in semiconductors either provide donor depths based on statistical evidence~\cite{Garleff_PRB_2008}, or only qualitatively locate dopants within a few nanometers region (of the order of 2.5 nm or more)~\cite{Mohiyaddin_Nanolett_2013}. The key metrological challenge in all of the ultra-scaled applications is the ability to determine the position of dopant atoms in the silicon crystal substrate with lattice-site precision, which will drastically transform our understanding at the most fundamental scale leading to devices with optimised functionalities. 

In this work, we present an atomically precise metrology and demonstrate the pinpointing of the position of subsurface phosphorous (P) and arsenic (As) dopants in silicon down to individual lattice sites. The technique involves low temperature STM imaging~\cite{Salfi_NatMat_2014} in conjunction with a fully quantum, large-volume treatment of the STM-dopant system. We demonstrate the metrology procedure using STM images obtained for several sub-surface P and As dopants. By comparing those images with a state-of-the-art multi-million-atom tight-binding framework~\cite{Usman_JPCM_2015, Rahman_PRL_2007} we determine the dopant lattice location which uniquely reproduces the normalised tunnelling current image with unprecedented accuracy over an 8$\times$8 nm$^2$ grid. The relatively large extent of the dopant electron wave function (Bohr radius $\sim$2 nm) as well as high-frequency contents occurring from the silicon valley interference processes~\cite{Salfi_NatMat_2014} result in the atomic level features in the STM image, which are highly sensitive to both the dopant position and the details of STM tip orbital involved. In all cases, we establish the STM tip state to be dominated by $d$-type tip orbital, consistent with the expectations of the tungsten tip~\cite{Chen_PRB_1990}. For each experimental STM image, a quantitative comparison to the images calculated at various dopant locations allows us to uniquely pinpoint the actual 3D lattice site position of the dopant with respect to the surface dimer rows for depths down to about 36 lattice planes ($\sim$5 nm). These results establish an STM-based metrology to determine the exact 3D position of dopants either in ultra-scaled nanoelectronic components, or in the construction of a quantum computer, with wider implications for the fabrication of atom-based devices more generally.  

\begin{figure*}
\centering
\singlespacing
\includegraphics[scale=0.1]{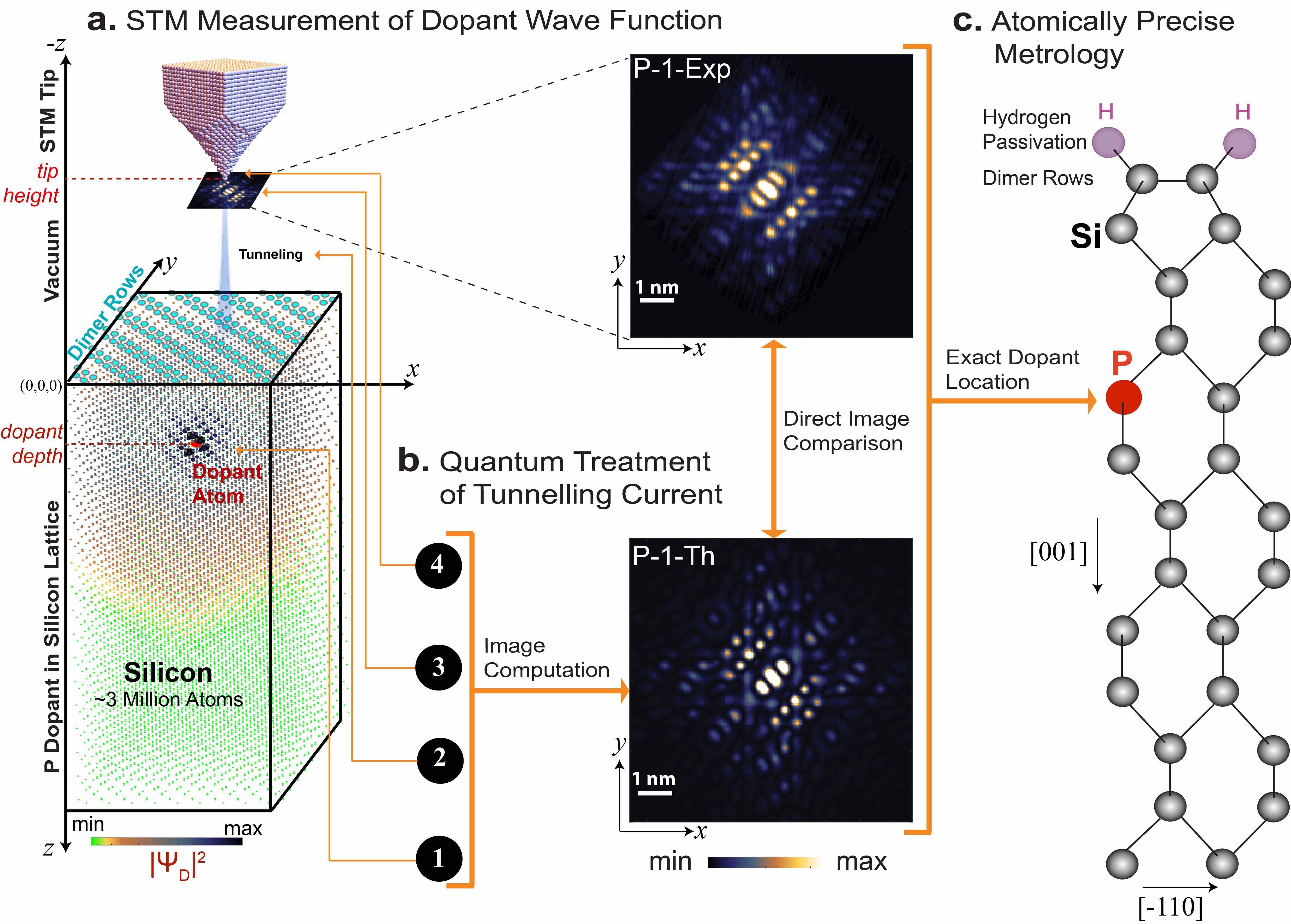}
\caption{\textbf{STM-based metrology for the exact position of sub-surface dopants in silicon:} \textbf{a.} Illustration of the STM setup used to measure atomically-resolved subsurface dopant images. The dopant charge density distribution ($|\Psi_ \textrm D|^2$) calculated from tight-binding simulation over a large-volume of the Si lattice is shown, which is color coded to highlight its variation as a function of the distance from the dopant atom. A P dopant is positioned a few nanometers below the surface ($z$=0) and is highlighted in red. The 2$\times$1 reconstruction results in the formation of dimer rows at the $z$=0 surface. The vertical shaded area in the vacuum region (between sample and STM tip) illustrates tunnelling of a single electron from the P dopant state in Si, through the vacuum region, to a single tungsten atom at the STM tip apex. The measured atomically-resolved image of the normalised tunnelling current is shown (P-1-Exp). The $x$ and $y$ axes represent [100] and [010] crystallographic directions, respectively. \textbf{b.} The theoretical framework involves the quantum mechanical treatment of the four linked components of the STM setup, as following: \textcolor{white}{\protect \circled{1}} Multi-million atom calculation of the dopant ground-state wave function with 2$\times$1 surface reconstruction ($\Psi_ \textrm D(r)$), \textcolor{white}{\protect \circled{2}} Wave function decay in the vacuum region ($\bigtriangledown^2 \Psi_ \textrm {D/T} - \kappa^2 \Psi_ \textrm {D/T}$)~\cite{Chen_PRB_1990}, \textcolor{white}{\protect \circled{3}} Bardeen tunnelling current ($I \propto \int_{\chi} ( \Psi_ \textrm T^* \nabla \Psi_ \textrm D - \Psi_ \textrm D \nabla \Psi_ \textrm T^* ). d\chi $)~\cite{Bardeen_PRL_1961}, and \textcolor{white}{\protect \circled{4}} STM tip state ($\Psi_ \textrm T(r)$) from the derivative rule~\cite{Chen_PRB_1990}. The final calculated image of the normalised tunnelling current (P-1-Th) is shown here. \textbf{c.} A direct quantitative comparison of the STM image and the images computed over a number of atomic positions leads to a unique identification of the subsurface dopant location in the silicon lattice.}
\label{fig:Figure1}
\end{figure*} 

Fig.~\ref{fig:Figure1} presents an overall sketch of our metrology technique. We first describe STM-donor measurement setup and the quantum mechanical treatment of the tunnelling current as shown in Fig.~\ref{fig:Figure1} (a) and (b). Our data set includes phosphorus (P-1 and P-2) and arsenic (As-1 and As-2) dopants. Although the image detail has a complex relationship to the quantum mechanics of the tip-orbital configuration, tip height above the sample, exact dopant position with regard to lattice plane depth, and lateral position relative to the surface dimers, we show explicitly these dependencies for a specific example (P-1). Fourier space analysis provides further evidence of the correct tip-orbital and surface reconstruction. By a direct quantitative (pixel-by-pixel) comparison with the experimental images (such as shown in Fig.~\ref{fig:Figure1} (c)) the precise dopant position is located with a high degree of confidence for all four members of the data set.
\\ \\
\noindent
\textcolor{blue}{
\textbf{\normalsize{STM measurement of dopant state}}} \\
The electronic wave functions of dopants buried in a Si crystal were spatially resolved by means of low temperature scanning tunnelling microscopy.~\cite{Salfi_NatMat_2014} Both P and As dopants were measured. The P dopants were incorporated in Si by $\rm PH_{3}$ dosing. A low P density was overgrown with 2.5 nm of Si by in-situ epitaxy~\cite{miwa2013} ($i.e.$ a target depth of 4.75$a_0$, where $a_0=0.5431$ nm is the Si lattice constant). A careful engineering of the growth temperature profile, including a 1 nm thin lock-in layer~\cite{Keizer_ACSNano_2015}, gives control on the depth of the P dopants by limiting segregation and diffusion while achieving the low surface roughness required for the STM measurements. The As dopants were found at random positions in a $\sim$10 nm thin thermally depleted region of highly-doped substrate. All samples consist of a degenerate Si substrate acting as an electron reservoir and an intrinsic region including a low concentration of P (or As). With this vertical sample geometry, the STM measurement falls into the single electron transport regime~\cite{Voisin_JPCM_2015}, and the wave function of electrons bound to dopants in this region was probed in real space at 5 K using a standard electrochemically etched tungsten tip. The experimental STM images in this study -- such as P-1-Exp shown in Fig.~\ref{fig:Figure1}(a) -- are normalised tunnelling current images, where the current $I(r)$ is proportional to $| \Im [\Psi_ \textrm D(r)]|^2$, with $\Im$ being a functional of the dopant wave function $\Psi_ \textrm D(r)$ determined by the STM tip orbital state~\cite{Chen_PRB_1990}. Full details of the experimental setup and the measurement procedures are provided in the section S1 of the supplementary information. 
\\ \\
\noindent
\textcolor{blue}{
\textbf{\normalsize{Quantum treatment of STM measurement}}} \\
The theoretical calculation of the measured STM images for the subsurface dopants is challenging as it requires an atomistic calculation of the dopant electron wave function over a volume comprising thousands of silicon atoms, including the surface reconstruction, and the coupling of this description to a quantum mechanical treatment of the tip-tunnelling process. We divide the development of a theoretical framework for this problem in four stages, as schematically shown in the Fig.~\ref{fig:Figure1}(b). First the calculation of the dopant wave function is performed by solving an \textit{sp$^3$d$^5$s$^*$} atomistic tight-binding Hamiltonian~\cite{Boykin_PRB_2004} over three million atoms including central-cell effects~\cite{Usman_JPCM_2015, Usman_PRB_2015, Rahman_PRL_2007}, which inherently incorporates the valley-orbit (VO) interaction in order to correctly capture the atomic fine detail for both P and As cases. As the dopants are close to the surface, the effects of dimer formation (2$\times$1 reconstruction)~\cite{Craig_SS_1990, Salfi_NatMat_2014} and hydrogen passivation~\cite{Lee_PRB_2004} are included in the tight-binding Hamiltonian (supplementary section S2).

\begin{figure}
\centering
\includegraphics[scale=0.1]{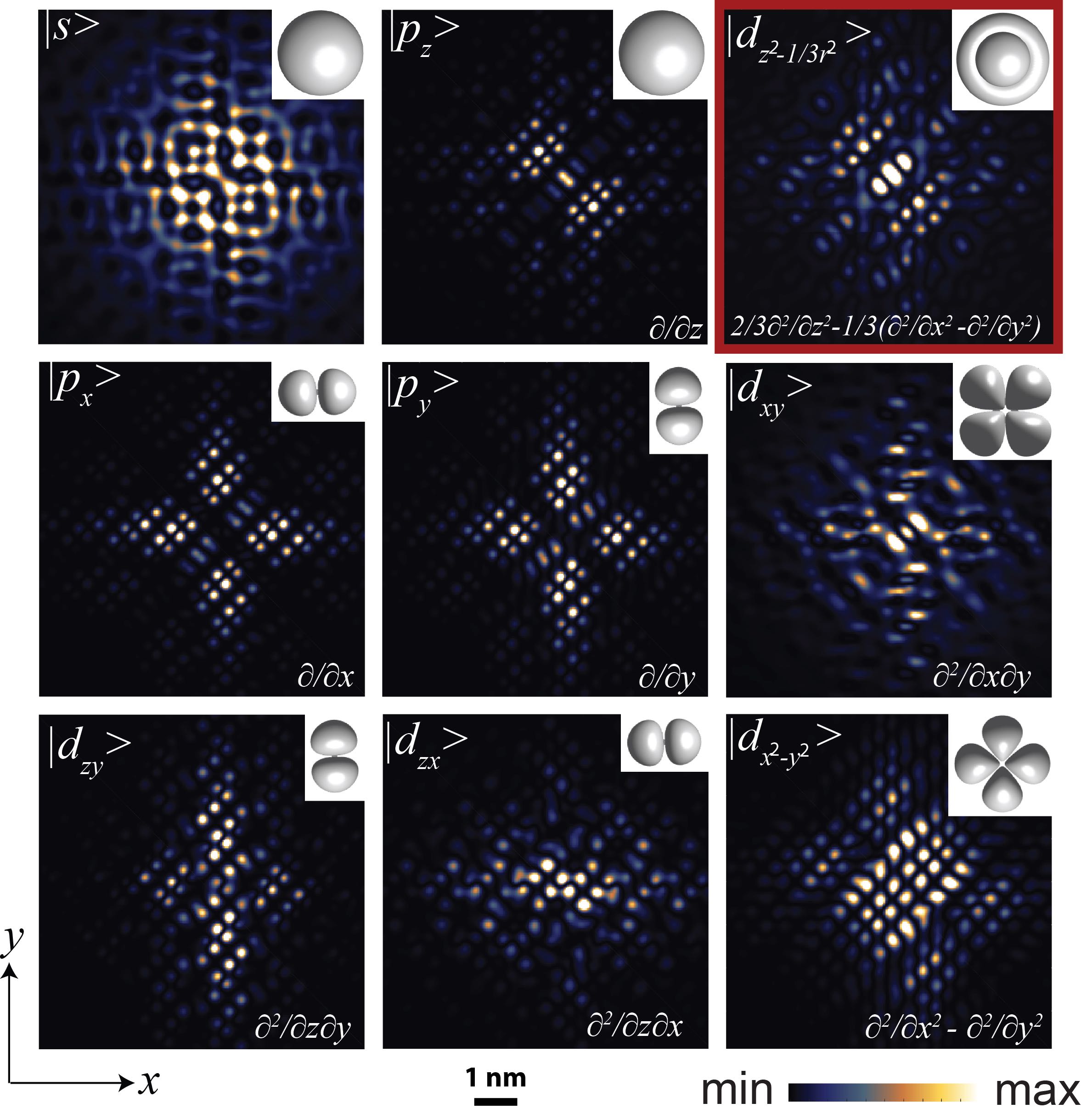}
\caption{\textbf{STM tip orbital dependence:} Theoretically computed $| \Im [\Psi_ \textrm D(r)]|^2$ images are shown for the P-1 dopant as a function of the individual orbitals in the STM tip state ($\Psi_ \textrm T$.) Each image is labelled by the corresponding tip orbital (top left of each image), the corresponding orbital symmetry in the $(x,y)$ plane (top right of each image), and the differential operator based on Chen's derivative rule~\cite{Chen_PRB_1990} applied to the sample wave function (bottom right of each image). Note that a $s$ tip-orbital will give an image directly proportional to the modulus squared of the wave function, without any derivative. The tip height is taken to be 2.5$\textrm{\AA}$ and the dopant depth is 4.75$a_0$. Each tip orbital produces a very different image of the dopant wave function, and only the $d_{z^2- \frac{1}{3}r^2}$ orbital image (highlighted by the red boundary) captures the symmetry of the atomic features present in the measured data P-1-Exp shown in Figure~\ref{fig:Figure1}(b). }
\label{fig:Figure2}
\end{figure} 

The STM measurements are based on the tunnelling of a single electron from the dopant ground state to the tip state through vacuum, so in the second and third stages of the theoretical framework Bardeen's tunnelling theory~\cite{Bardeen_PRL_1961} is coupled with the tight-binding dopant wave function to facilitate the computation of the tunnelling current and hence the image corresponding to the $| \Im [\Psi_ \textrm D(r)]|^2$ (supplementary section S3). The vacuum decay of the tight-binding wave function is based on the Slater orbital real-space dependence~\cite{Slater_PR_1930}, which satisfies the vacuum Schr\"{o}dinger equation~\cite{Chen_PRB_1990}. The exponential decay of the tunnelling current in the $z$-direction ($I(z+dz)$=$I(z)e^{-2 \kappa dz}$) is characterised by a decay constant $\kappa$. Our calculated values of $\kappa$ in the range of 0.013-0.016 pm$^{-1}$ for various dopant positions are in good agreement with the measured value of $\sim$0.01 pm$^{-1}$ by controlled variation of STM tip position. This validates the evanescent decay of the tight-binding wave function through the vacuum to the tip apex. 

\begin{figure*}
\centering
\includegraphics[scale=0.2]{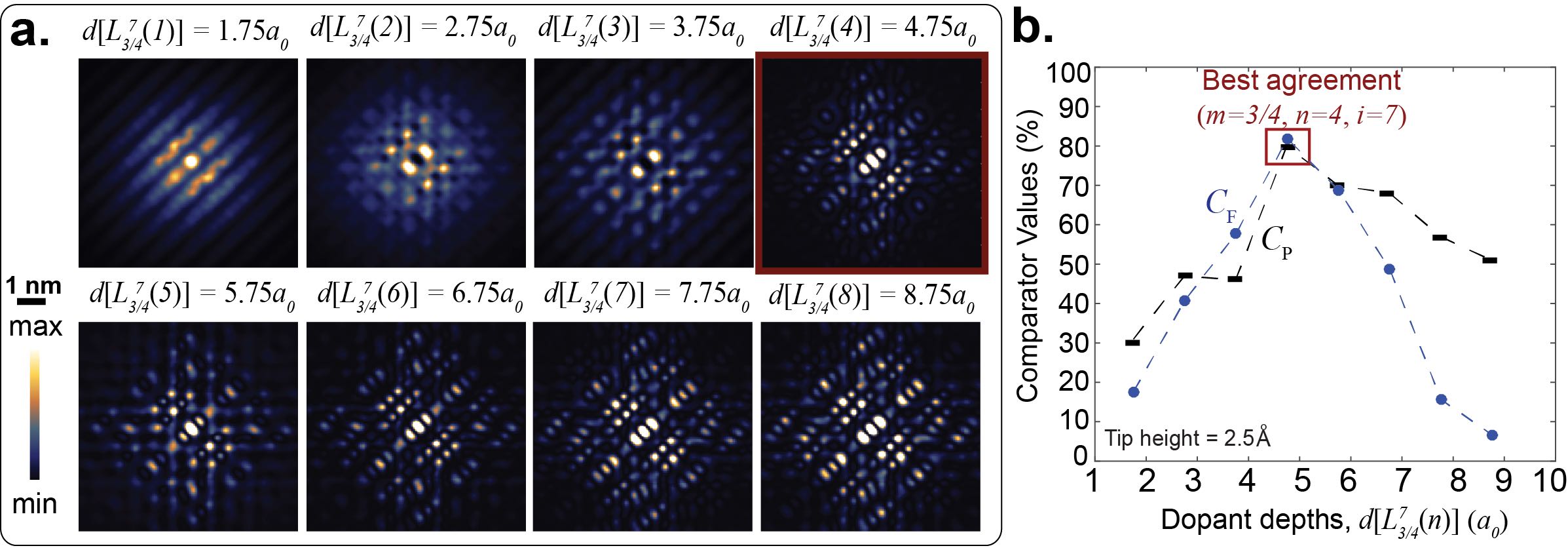}
\caption{\textbf{Unique lattice site assignment for P-1 dopant by depth analysis:} \textbf{a.} To identify the exact location of the measured P-1-Exp image, theoretically computed images are shown as a function of plane depth, $n$, once the $L_{3/4}^7(n)$ location group has been identified from the image symmetry (details are in Fig. S4 of the supplementary section). The computed image at 4.75$a_0$ depth (highlighted with the red colour border), corresponding to $n$=4, exhibits the best quantitative agreement with the measured P-1-Exp image as detrained by the comparator peaks. \textbf{b.} Plot of the pixel-by-pixel $C_{\rm P}$ and feature-by-feature $C_{\rm F}$ comparators between the P-1-Exp image and the theory images displayed in part (a).  Best agreement with the P-1-Exp image is uniquely identified at the dopant depth of 4.75$a_0$ ($m=3/4, n=4, i=7$) by joint peaks of the both comparator parameters. The corresponding depth plane analysis for the other dopants in the set (P-2-Exp, As-1-Exp, and As-2-Exp) are shown in Supplementary Figure S9.}
\label{fig:Figure3}
\end{figure*} 

The fourth stage of the dopant image calculation involves determination of the electronic state of the STM tip, which is not known $a$ $priori$. To comprehensively explore the role of tip orbitals and to definitively determine the tip electronic state responsible for the tunnelling current in our measurements, we start with a general tip state, described as $\Psi_ \textrm T$ = $\sum_{\beta=1}^{9} A_{\beta} \phi_{\beta}^ \textrm {T}$, where $\phi_{\beta}^ \textrm {T}$ is $s, p_x, p_y, p_z, d_{xy}, d_{zy}, d_{zx}, d_{x^2-y^2}, d_{z^2- \frac{1}{3}r^2}$ orbitals for the $\beta$ = 1 to 9, respectively. The contribution from each tip orbital is based on Chen's derivative rule~\cite{Chen_PRB_1990}, in which the tunnelling current is proportional to derivative (or the sum of derivatives) of the sample wave function computed at the tip location (supplementary section S3). Through a systematic and extensive exploration of the parameter space of $\Psi_ \textrm T$ state (by varying $A_{\beta}$), we find a tip orbital configuration that leads to the best agreement with the experimental measurement. This procedure necessarily involves a rigorous search through physical parameters such as dopant depth, dopant position underneath dimer rows, and tip height above the Si surface. Finally the quantitative comparison of the measured and calculated images is performed by defining two comparator parameters: pixel-by-pixel difference ($C_{ \rm P}$) and feature-by-feature correlation ($C_{ \rm F}$) of images (supplementary section S3). \\   

\noindent
\textcolor{blue}{
\textbf{\normalsize{STM tip state and surface reconstruction}} } \\
Fig.~\ref{fig:Figure1} (b) plots the final (normalised) theoretical STM image for the phosphorus dopant P-1. The final converged images for other three dopants (P-2, As-1, and As-2) are shown in the Supplementary Fig. S7, along with the corresponding STM measurements, all exhibiting excellent agreement between theory and experiment. In the computation of the images the unknown physical quantities related to the STM measurement setup form a large parameter space to explore. In general, these include the tip orbital configuration ($\Psi_ \textrm T$), height of the tip above the $z$=0 surface, depth of the dopant atom below the $z$=0 surface, and position of the dopant atom with respect to the surface dimer rows. We systematically tackle this complex problem space to obtain the optimal agreement between theory and experiment. We first focus on how varying those parameters affect the images, using the dopant P-1 as example.

By employing the generalised tip state and optimising the tip orbital weights, $A_{\beta}$, we can study the dopant images as a function of the tip orbital configurations. To illustrate the effect of tip-orbital on the calculated images, in Fig.~\ref{fig:Figure2}, we plot the theoretical images for the dopant P-1 at a depth of 4.75$a_0$ and a tip height of 2.5$\textrm{\AA}$, as a function of the individual tip orbitals. Evidently each tip orbital captures a very different symmetry of the same dopant wave function. With reference to the experimental image (P-1-Exp) in Fig.~\ref{fig:Figure1}(b) the character of the image is clearly dominated by the $\Psi_ \textrm T$ = $d_{z^2- \frac{1}{3}r^2}$ tip state. We consistently find by far (more than 90\%) the dominant contribution from the $d_{z^2- \frac{1}{3}r^2}$ orbital in the tip state reproducing all of the measured images~\footnote{More than twenty measured P and As images -- including the four images P-1-Exp, P-2-Exp, As-1-Exp, and As-2-Exp presented in this work -- were studied and in all cases we found image structures clearly dominated by $d_{z^2- \frac{1}{3}r^2}$ orbital character in the tip state.} with the best quantitative agreement (see Supplementary Fig. S7). This is consistent with the existing density of states based arguments~\cite{Chen_PRB_1990, Chaika_EPL_2010, Teobaldi_PRB_2012} that $d$-type orbitals should dominantly contribute in the tunnelling currents involving STM tips made up of transition metal elements. The quantitative agreement between our measurement and theory therefore provides a concrete verification of the dominant $d$-wave tunnelling -- in our case major contribution is from the $d_{z^2- \frac{1}{3}r^2}$ orbital. We infer this to the presence of high-frequency components arising from the valleys of Si-dopant system and the unusually large spatial extent of the wave function (compared to the atomic sized features) of dopants below the surface, which are highly sensitive to the precise tip-orbital configuration. In Supplementary Fig. S5, we compare the Fourier Transform spectrum (FTS) of the measured and calculated images for the P-1 dopant. The excellent agreement confirms the dominant role of $d_{z^2- \frac{1}{3}r^2}$ orbital in tunnelling current, and the 2$\times$1 surface reconstruction scheme. 
\begin{figure*}
\centering
\includegraphics[scale=0.125]{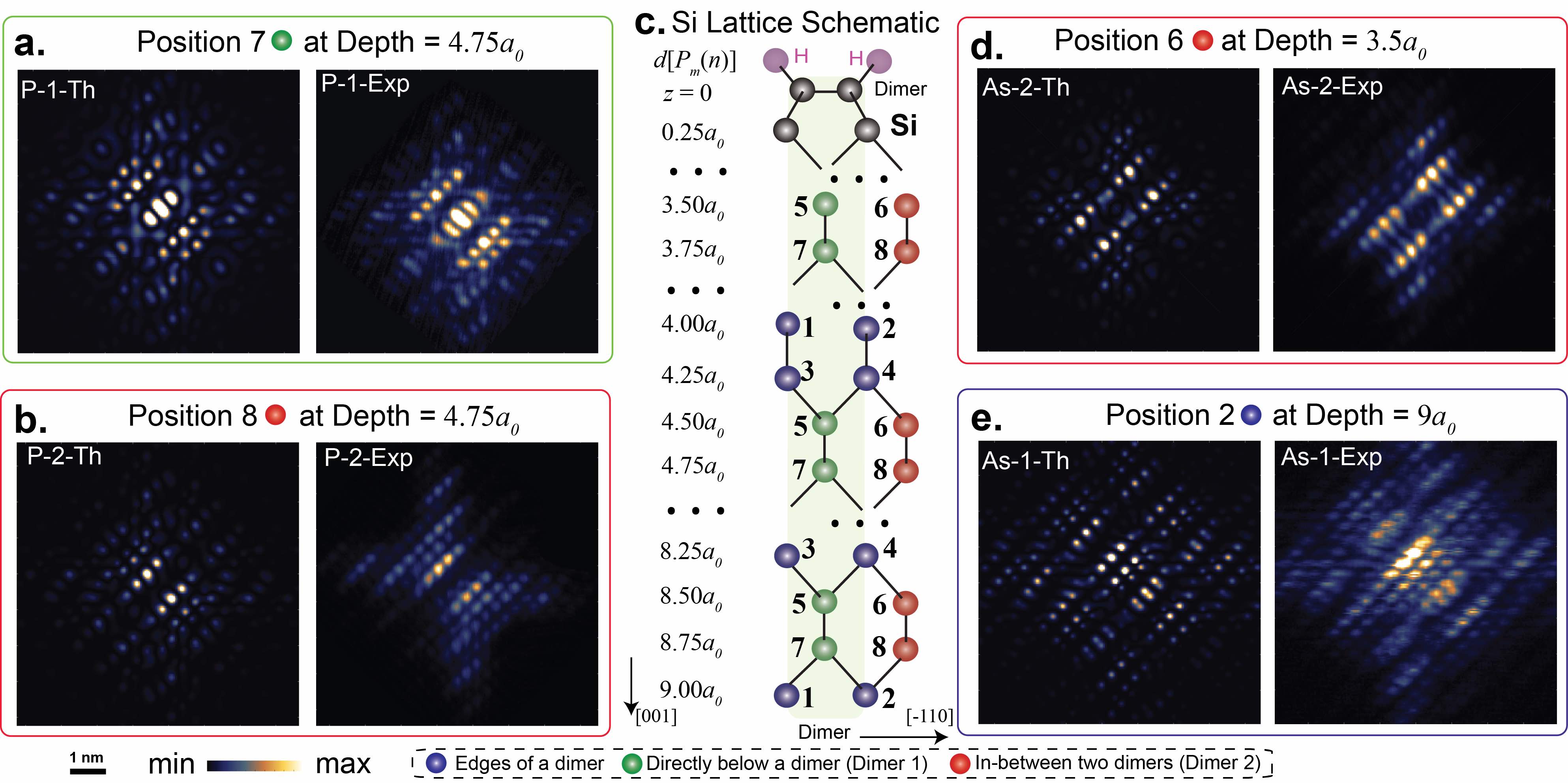}
\caption{ \textbf{Pinpointing the dopant location with single-lattice site precision:} \textbf{a. and b.} Theoretically computed and experimentally measured images (equivalently normalised) are shown for the dopants P-1 and P-2 at the final determined lattice site positions $L^7_{3/4}(4)$ and $L^8_{3/4}(4)$ (depth=4.75$a_0$), respectively. Despite being at the same depth from the $z$=0 surface, the symmetries of the P-1 and P-2 images reflect the different positions relative to the surface dimer rows (directly underneath the dimer and in-between the two dimer rows, respectively). The location of the dopants is therefore uniquely determined in depth and with respect to the surface dimer positions. \textbf{c.} Illustration of the relevant silicon lattice sites in the silicon crystal structure. The topmost surface of the silicon is hydrogen passivated (hydrogen atoms are shown by purple colour). At the $z$=0 surface, dimer rows are shown along the direction perpendicular to the page, as indicated by the dark color sites. The three categories of the lattice positions are highlighted by blue, green, and red coloured atoms with respect to the dimer rows: green=positions directly below the dimer rows, red=positions in-between two dimer rows, and blue=positions at the edges of a dimer row. Due to symmetry properties, green and red positions would correspond to unique patterns of features in the images, whereas the blue positions simply lead to 180$^o$ rotation of the image features. \textbf{d. and e.} Theoretically computed and experimentally measured images (equivalently normalised) are shown for the dopants As-1 and As-2 at the final determined lattice site positions $L^6_{1/2}(3)$ (depth=3.5$a_0$) and $L^2_{0}(9)$ (depth=9$a_0$), respectively.}
\label{fig:Figure4}
\end{figure*}
\\ \\
\noindent
\textcolor{blue}{
\textbf{\normalsize{Pinpointing the exact dopant location}} } \\
After validation of tip orbital and surface reconstruction, next we demonstrate how the depth and the lateral position of the dopant with respect to the surface dimers has a significant effect on the image symmetry, and ultimately permit the identification of the actual dopant lattice site. In Supplementary Fig. S8 we plot a schematic diagram of the Si crystal, establishing our notation regarding the possible donor positions below the surface. Along the [001] direction, each lattice position corresponds to a unique depth and therefore should lead to a distinct image. A closer examination of the dopant images uncovers subtle symmetry effects, which repeat for every fourth plane. Therefore the Si lattice planes are divided into four plane groups ($PG_m$, where $m \in$ \{0, 1/4, 1/2, 3/4\}). Each plane group is a set of planes $P_{m}(n)$, whose depths from the $z$=0 surface are given by: $d[P_{m}(n)]$=($m$+$n$)$a_0$, where $n$=0,1,2,3,... Note that $(m,n)$=(0,0) corresponds to the $z$=0 surface. Considering the periodicity of the dimers along the [-110] direction, two possible dopant locations $L_{m}^i(n)$ per plane $P_{m}(n)$ are defined by $i=8m+1$ and $i=8m+2$ which repeat periodically in the lateral direction. The dopant locations $L_{m}^i(n)$ are color coded in the Supplementary Fig. S8 with respect to the dimer rows -- the eight positions marked as $i$=1 to 8 periodically repeat in all crystal directions. The positions $i$=1\&2 and 3\&4 (colored blue) are on the edges of the dimer rows and therefore are equivalent: at a given depth $n$, dopants at locations 1 and 3 will lead to the same image features as locations 2 and 4 respectively, with a 180$^o$ rotation. The allocation of exact dopant site (1 or 2, 3 or 4) is based on overlying dimer positions on top of the image and finding dopant location on either edge of the dimers. Overall the classification with respect to the unique patterns of image features leads to six possible distinguishable dopant location-groups in the Si lattice given by: $L_{0}^{1,2}(n)$, $L_{1/4}^{3,4}(n)$, $L_{1/2}^5(n)$, $L_{1/2}^6(n)$, $L_{3/4}^7(n)$, and $L_{3/4}^8(n)$. For an experimental STM image, our metrology technique first associates it with one of the six location-groups (by determining $m$ and $i$), followed by quantitative depth comparison of images within that group (by varying $n$).

The assignment of a location group to a measured STM image is based on a systematic procedure derived from careful analysis of the symmetries of the image features. The details of the procedure are described in supplementary information section S4, along with its application to the four measured STM images. After determining a location group for the measured image -- for example $L_{3/4}^7(n)$ location-group for the P-1-Exp image -- next we quantitatively compare the computed images within the allocated location group with the measured image by varying the plane depth, $n$. Fig.~\ref{fig:Figure3}(a) shows the analysis for the P-1-Exp image where $n$ is varied from 1 to 8, and the corresponding comparator graphs ($C_{\rm P}$ and $C_{\rm F}$) between each theory and P-1-Exp image are plotted in Fig.~\ref{fig:Figure3}(b). The highest comparator values ($C_{\rm P}$=80\% and $C_{\rm F}$=82\%) uniquely locate the P dopant at position $L_{3/4}^7(4)$ (4.75$a_0$ depth) as plotted in Fig.~\ref{fig:Figure4}(a). In order to include the effect of tip-height variation in the theoretical calculation and/or instabilities of tip-height in the measurements, we changed the 2.5$\textrm{\AA}$ tip-height by $\pm$0.125$\textrm{\AA}$ and recomputed the values of $C_{\rm P}$ and $C_{\rm F}$ for $n$=4. This $\pm$5\% tip-height variation introduces small changes in the comparator values (less than 2\%), indicating that such effect is not expected to influence the optimum agreement between the theory and experiment for the dopant location of $L_{3/4}^7(4)$, computed with the $d_{z^2- \frac{1}{3}r^2}$-wave tip.  

The application of the procedure to the other dopants in the data set, P-2-Exp, As-1-Exp, and As-2-Exp, gives similar behaviour in their respective comparators (Supplementary Fig. S9), which allows us to uniquely identify their positions. For the whole data set the final results for positions and computed images are shown in Fig.~\ref{fig:Figure4}. We should emphasise here that the existence of the peak in comparator graphs is more meaningful than the value itself, as the measurements include some invariably present perturbations to the dopant wave function that have not been considered in the model. In general, these perturbations include applied electrostatic potentials from the STM tip itself, and background electrostatic disorder that is an aggregate of charge density disorder in the buried reservoir and nearby ($>10$ nm away) dangling bonds. As discussed in Refs.~\onlinecite{Salfi_NatMat_2014,Voisin_JPCM_2015}, we can control the tip-induced band bending during dopant state measurements, and as desired in the present work, we minimize the potential applied by the STM tip. The residual fields are typically on the order of 2 MV/m~\cite{Voisin_JPCM_2015}, and thus, are much smaller than those predicted to hybridize the dopant wave function with an interface-well wave function~\cite{Rahman_PRB_2011}. Furthermore, Bardeen's first-order tunnelling approach~\cite{Bardeen_PRL_1961} does not take into account the higher-order tip-induced modifications to the wave function. It is remarkable that despite these simplifications in the model, both $C_{\rm P}$ and $C_{\rm F}$ comparator peaks coexist at the same dopant location with their values exceeding 60\% and all of the measured features one-to-one match with computed features (Supplementary Fig. S7), enabling the unambiguous pinpointing of the exact dopant lattice site position in each of the four data sets.

The comparison for the deepest dopant As-2 (Fig.~\ref{fig:Figure4} (d)) shows that the procedure is reaching its limit at a depth of 9$a_0$, corresponding to the 36th lattice planes below the $z$=0 surface. We also remark that a comparison of the computed P and As images revealed large differences at small dopant depths (3$a_0$ or less), which drastically decrease as the dopant positions get deeper. This is expected as the P and As wave functions primarily differ by the central-cell effects which are tightly confined closer to the dopant atoms. A comparison of (a) and (b) in Fig.~\ref{fig:Figure4} highlights the effect of the surface reconstruction, resulting in drastically different spatial symmetries of the images despite the corresponding P dopants being at the same 4.75$a_0$ depth from the $z$=0 surface. On the contrary, the FTS of these two images shown in the Supplementary Fig. S6 (a) and (b) do not differ in any great detail. This implies that whilst the position of dopant with respect to dimers is crucial in determining the symmetry of STM-measured images; it has very little impact on the actual dopant wave function and its underlying valley configuration~\footnote{The impact of the lateral surface strain due to the 2$\times$1 reconstruction is found to be very weak on the valley configurations, which is primarily governed by the depth of the dopant along the [001] axis. This is confirmed by calculating valley populations for the P-1 and P-2 dopants, which only differ by less than 0.1\%, compared to $\sim$1.25\% change if P-1 at 4.75$a_0$ depth is compared to a P dopant at 4.5$a_0$ depth.}. Finally, it is interesting to note that the metrology of both P dopants reveal their depths at 4.75$a_0$, which coincide with the incorporated target depth. Indeed, this metrology now allows assays to be performed, which could determine whether this is really indicative of low segregation during the fabrication process.
\\ \\
\noindent
\textcolor{blue}{
\textbf{\normalsize{Summary and outlook}} } \\
STM measurements in conjunction with large-scale quantum simulations provide an atomic-precision metrology procedure for finding the exact locations of sub-surface group V dopants in silicon. The large Bohr radius of the dopant electron (compared to the silicon lattice constant) provides a rich variety of atomic-sized features in the STM images, the symmetry and details of which are highly sensitive to the absolute position of the dopant below the surface. In the atom-by-atom fabrication of nanoscale electronic devices more generally, the determination of the precise positions of countable dopants is critical to device performance and is expected to lead to optimisation strategies for fabrication and characterisation. The critical role of the STM tip state in our measurements is established, which is necessary to understand the both real and Fourier space spectra of dopant wave function from STM images (supplementary section S5). The high-level agreement between theory and experiment achieved here for the normalised tunnelling current images can be pushed further to understand hitherto hidden subtle effects such as central-cell corrections at the dopant site, local strain conditions, surface effects, valley interference and non-static screening of the dopant electron in the silicon crystal. Precision metrology of dopant locations in the fabricated closely-spaced dopant pairs~\cite{Salfi_arXiv_2015} will enable probing of multi-particle physics, and provide an ideal test-bed to answer fundamental questions pertaining molecular physics. For quantum computer architectures~\cite{Hill_science_2015}, with exact relative dopant qubit distances known during the fabrication stage, highly precise quantum logic gates can be constructed -- a fundamental ingredient for quantum error correction and scale-up.
\\ \\
\noindent
\textcolor{black}{
\textbf{Acknowledgements:}} This work is funded by the ARC Center of Excellence for Quantum Computation and Communication Technology (CE1100001027), and in part by the U.S. Army Research Office (W911NF-08-1-0527).  MYS acknowledges an ARC Laureate Fellowship. Computational resources are acknowledged from NCN/Nanohub. MU acknowledges useful discussions with Charles Hill and Viktor Perunicic. 
\\ \\
\noindent
\textcolor{black}{
\textbf{Author contributions:}} LCLH and MU formulated the theoretical framework for the metrology scheme, including tight-binding calculations of the STM images with generic tip orbitals with input from JS. MU performed the theoretical calculations. JB, JS, BV, MYS and SR designed and conducted the STM measurements. All authors provided input in writing of the manuscript.
\\ \\
\noindent
\textcolor{black}{
\textbf{Additional information:}} The accompanied supplementary information document provides additional details on methods, quantitative comparison of measured and computed STM images, and their Fourier analysis. 
\\ \\
\noindent
\textcolor{black}{
\textbf{Competing financial interests:}} The authors declare no competing financial interests. 

\newpage

\begin{center}
\textbf{{{\LARGE \underline{Supplementary Information}}}}
\end{center}

\begin{center}
\textcolor{blue}{
\textbf{S1. Experimental Procedures} } \\
\end{center}

\noindent
\textbf{Sample Preparation:}
Two different types of samples, with either P or As dopants, were prepared in UHV within a low temperature scanning tunnelling microscopy (STM) system. Samples with As dopants were prepared by flash annealing a $0.003$ ohm-cm As-doped wafer three times at approximately 1050 $^\circ$C for a total of 30 seconds. In these conditions, approximately 10 nm from the Si surface are depleted, leaving an As concentration of $\approx 2\times10^{17}$ $cm^{-3}$.~\cite{Salfi_NatMat_2014} Consequently, As dopants measured in these samples were found at random depths in a $\sim$10 nm thin thermally depleted Si layer below the surface. Samples with P dopants were prepared by incorporating P in Si by a sub-monolayer phosphine ($\rm PH_{3}$) dosing of the previously flash annealed Si substrate~\cite{miwa2013}. A sheet density of $5\times10^{11}$ $cm^{-2}$ P donors is overgrown epitaxially by $\sim$2.5 nm (4.75$a_{0}$, where $a_{0}$ is Si lattice constant) of Si. The first nm is a lock-in layer grown at room temperature~\cite{Keizer_ACSNano_2015} and growth parameters, like temperature and fluxes, were chosen to achieve minimal segregation and diffusion. Consequently the depth of P donors is controlled. The 2$\times$1 reconstructed Si (001) surface of all samples was passivated with hydrogen (H). The H termination was carried out at $T=340$ $^\circ$C for approximately 10 min under a flux of atomic hydrogen produced by a thermal cracker, in a chamber with a $10^{-7}$ mbar pressure of molecular hydrogen. Both P and As doped samples consisting of a degenerate Si substrate acting as an electron reservoir and an intrinsic region including a low concentration of donors, the STM measurement falls into the single electron transport regime described in Ref.~\onlinecite{Voisin_JPCM_2015}. A schematic of this vertical structure and the STM junction is shown in Fig. 1(a) of the main text. 

\noindent
\textbf{Spatially resolved transport measurements:}
The samples were measured by STM and spectroscopy at 4.2 K in UHV using an electrochemically etched tungsten (W) tip. We observed spatially localized states in constant current images at a nominal sample bias $U=-1.25$ V and tunnel current 100 pA. The spatially localized features resolved into a series of single-electron transport peaks in bias-resolved tunnelling spectroscopy~\cite{mol2013, Salfi_NatMat_2014, Voisin_JPCM_2015}. The lowest-energy state of a donor was measured with high resolution in real space using an unconventional two-pass scheme, where the topography was first measured at $U=-1.45$ V. Replaying the topography in the second pass, we measured the current in open-loop mode, typically at a bias $U=-0.8$ V where only the neutral donor state is in the bias window, and constant tip offset of 200 pm towards the sample to enhance the tip-donor coupling. Owing to complex valley quantum interference processes observed upon Fourier transformation of the measured state, and supported by the state's proximity to the empirically determined flat-band condition, the lowest energy state was attributed to transport through the ground state of the neutral donor charge configuration~\cite{Salfi_NatMat_2014}. The broadening of the lowest energy peak is always essentially independent of tip height, and the transport line shape matches what is expected for broadening due to coupling to a thermally broadened Fermi sea~\cite{averin1991} at liquid Helium temperature. We attribute this to weak tunnel coupling of the donor to a buried Fermi sea, as expected for flash annealed Si substrates~\cite{pitters2012}.

\noindent
\textbf{Donor depth measurement for As-2 data:}
The depth of a dopant beneath a semiconductor surface can be estimated by comparing measurements of the spatial perturbation of a feature such as a band edge by the ionized dopant, based on comparison of spectroscopic data to an electrostatic model. One such model, first used for dopants in GaAs~\cite{lee2010, teichmann2008}, is based on a dielectric screened Coulomb potential with a discontinuity at the semiconductor/vacuum interface. This model has recently been applied to single acceptors~\cite{mol2013} and donors~\cite{Salfi_NatMat_2014} in Si. Following to reference~\onlinecite{Salfi_NatMat_2014}, we have employed this method to estimate the depth of the As-2-Exp donor. To do so, we measured the spatial variation of the onset of the conduction band edge at a positive sample bias $U>0$, well above the flat-band condition $U\approx-0.8$ V, where the neutral donor state is above the Fermi energy of the highly doped reservoir. The sample bias $U_C$ required to obtain a tunnel current of $0.1$ pA was spatially mapped in two dimensions $(x,y)$ as a function of position, and fit to a Coulomb potential for the donor As-2-Exp. We obtained a depth $z_0= (5\pm1)a_0$ nm, which in spite of simplistic assumptions, including static dielectric constant for Si material and dielectric averaging across the silicon-vacuum interface, is reasonably close to the 3.5$a_0$ depth determined by our metrology.

\noindent
\textbf{Electric field measurement:}
This section presents experimental procedures to extract quantitative parameters such as tip-height variations and electric fields at the donor site, following references~\onlinecite{Salfi_NatMat_2014} and \onlinecite{Voisin_JPCM_2015}. The electric field experienced by the donor bound electron during the measurement is determined by measurements of the tunnelling spectrum taken with different tip heights. We compare these measurements to a one-dimensional model for electrostatics. At the flat-band condition, there is no electric field present in the vacuum, and following simple electrostatic arguments, the bias required to bring the state into resonance with the reservoir is independent of tip height. However, when there is a non-zero electric field in vacuum, a small change in bias is required to bring the state in resonance, when the tip height is changed. From the required bias and change in tip height we can quantitatively extract the field in vacuum, and using a simple model for dielectric screening, the electric field in the silicon. Typical values of electric fields experienced by donors, when imaged in resonant conditions, are less than 2 MV/m in magnitude.

Tip height variations discussed in this section were calibrated to the height of step edges on the Si (001) surface, which is a quarter of the cubic lattice constant of silicon. However the absolute tip height cannot be quantitatively estimated as it depends on the exact barrier and tip shapes, and the resulting coupling with the donor evanescent wave function, quantities which cannot be precisely determined at such small length scales. Offsets of 200 to 300 pm are typically used during the second pass. Larger offsets towards the surface can bring the tip in contact with the surface.

\begin{center}
\textcolor{blue}{
\textbf{S2. Multi-million Atom Tight-binding Calculations of dopant Wave functions} } \\
\end{center}

The atomistic simulations of electronic energies and states for a dopant in silicon are performed by solving an \textit{sp$^3$d$^5$s$^*$} tight-binding Hamiltonian. The \textit{sp$^3$d$^5$s$^*$} tight-binding parameters for the Si material are obtained from Boykin \textit{et al}.~\cite{Boykin_PRB_2004}, which have been optimised to accurately reproduce the Si bulk band structure. The dopant atoms (P and As) are represented by a Coulomb potential screened by non-static dielectric of the Si~\cite{Usman_JPCM_2015, Usman_PRB_2015}, truncated to U($r_0$)=U$_0$ at the dopant site ($r_0$). Here U$_0$ is an adjustable parameter that represents the central-cell correction and has been designed to accurately match the ground state binding energies of the P and As dopants as measured in the experiment~\cite{Ahmed_Enc_2009, Rahman_PRL_2007, Rahman_PRB_2009}. The size of the simulation domain (Si box around the dopant) is chosen as 40 nm$^3$, consisting of roughly 3 million atoms, with closed boundary conditions in all three spatial dimensions. The effect of Hydrogen passivation on the surface atoms is implemented in accordance with our published recipe~\cite{Lee_PRB_2004}, which shifts the energies of the dangling bonds to avoid any spurious states in the energy range of interest. The multi-million-atom real-space Hamiltonian is solved by a parallel Lanczos algorithm to calculate the single-particle energies and wave functions of the dopant atom. The tight-binding Hamiltonian is implemented within the framework of NEMO-3D~\cite{Klimeck_1, Klimeck_2}.

In our experiments, the (001) sample surface consists of dimer rows of Si atoms. We have incorporated this effect in our atomistic theory by implementing 2$\times$1 surface reconstruction scheme, in which the surface silicon atoms are displaced in accordance with the published studies.~\cite{Craig_SS_1990, Salfi_NatMat_2014} The impact of the surface strain due to the 2$\times$1 reconstruction is included in the tight-binding Hamiltonian by a generalization of the Harrison's scaling law~\cite{Boykin_PRB_2004}, where the inter-atomic interaction energies are modified with the strained bond length $d$ as $(\frac{d_0}{d})^{\eta}$, where $d_0$ is the unperturbed bond-length of Si lattice and $\eta$ is a scaling parameter whose magnitude depends on the type of the interaction being considered and is fitted to obtain hydrostatic deformation potentials. These models have previously demonstrated excellent agreements with the experimental findings~\cite{Klimeck_1, Klimeck_2}. It is emphasized here that the atomistic nature of our model has allowed a straightforward implementation of the surface reconstruction effect in our simulations, which would be inaccessible in continuum models based on effective-mass theories~\citep{Koiller_PRB_2002}.  

\begin{center}
\textcolor{blue}{
\textbf{S3. Calculation of Spatially-resolved STM Images} } \\
\end{center}

The calculation of the STM images is implemented by coupling the Bardeen's tunnelling theory~\cite{Bardeen_PRL_1961} and the derivative rule formulated by Chen~\cite{Chen_PRB_1990} with our tight-binding wave function. In the tunnelling regime, the relationship between the applied bias (V) on the STM tip and the tunnelling current (I) is provided by the Bardeen's formula:

\begin{equation}\label{IV}
{I(V)} = \frac{2 \pi e}{\hbar} \sum_{\mu \nu} (1 - f(E_\nu + eV)) |M_ \textrm {DT}|^2 \times \delta(E_\mu - E_\nu -eV)
\end{equation}
\noindent
where $e$ is the electronic charge, $\hbar$ is the reduced Planck's constant, $f$ is the Fermi distribution function, and $M_ \textrm {DT}$ is the tunnelling matrix element between the single electron states of the dopant (denoted by the subscript $\textrm D$) and of the STM tip (denoted by the subscript $\textrm T$). As derived by Chen in Ref.~\onlinecite{Chen_PRB_1990} that the tunnelling matrix element, for all the cases related to STM measurements, can be reduced to a much simpler surface integral solved on a separation surface $\chi$ arbitrarily chosen at a point in-between the sample and STM tip. Therefore,

\begin{equation}\label{tm}
{M_ \textrm {DT}} = \frac{\hbar^2}{2 m_e} \int_{\chi} ( \Psi_ \textrm T^* \nabla \Psi_ \textrm D - \Psi_ \textrm D \nabla \Psi_ \textrm T^* ). d\chi 
\end{equation}
\noindent
where $\Psi_ \textrm D$ is the single electron state of the sample (P or As dopant in Si), $\Psi_ \textrm T$ is the state of the single atom at the apex of the STM tip, and $d\chi$ is an element on the separation surface $\chi$. 

In our calculation of the STM images, we follow Chen's approach~\cite{Chen_PRB_1990}, which reduces equation~\ref{tm} to a very simple derivative rule where the tunnelling matrix element is simply proportional to a functional of the sample wave function computed at the tip location, $r_0$ (for $\chi$ assumed to be at the apex of the STM tip):

\begin{equation}\label{func}
{M_ \textrm {DT}}  \varpropto \Im [\Psi_ \textrm D (r)] 
\end{equation}
\noindent
where the functional of the wave function, $\Im [\Psi_ \textrm D (r)]$, is defined as a derivative (or the sum of derivatives) of the sample wave function -- the direction and the dimensions of the derivatives depend on the orbital composition of the STM tip state. The definitions of $\Im [\Psi_ \textrm D (r)]$ with respect to the tip state orbitals are provided in table~\ref{tab:table1}. 

The determination of the tip state is the hardest challenge in any theoretical calculation of STM tunnelling current as the tip electronic structure is not known $a$ $priori$. Much of the published literature~\cite{Bozkurt_PRB_2013, Salfi_NatMat_2014, Lemay_Nat_2001} is based on the model first proposed by Tersoff and Hamann~\cite{Tersoff_PRB_1985}, where the $s$-type tip orbital assumption drastically reduced the computational burden by leaving the measured tunnelling current proportional to the sample wave function charge density calculated at the tip location r$_0$, $i.e.$ $\Im [\Psi_ \textrm D (r)]$= $\Psi_ \textrm D$($r$). However, Chen~\cite{Chen_PRB_1990} proposed -- based on density of state argument -- that for the STM tips made up of transition metal elements, the tunnelling current should be dominant by the $d$-type orbitals in the tip state. Recent studies~\cite{Gross_PRL_2011, Teobaldi_PRB_2012, Mandi_PRB_2015, Li_NanoRes_2015, Chaika_SciRep_2014} explored the role of the higher orbitals in tip state based on Chen's proposal. We have also coupled the same derivative rule with our TB wave function. In order to comprehensively explore the role of tip orbitals and to quantitatively resolve the tip state in our measurements, we start with a generalized tip state, described as $\Psi_ \textrm T$ = $\sum_{\beta=1}^{9} A_{\beta} \phi_{\beta}^ \textrm {T}$, where $\phi_{\beta}^ \textrm {T}$ is $s, p_x, p_y, p_z, d_{xy}, d_{zy}, d_{zx}, d_{x^2-y^2}, d_{z^2- \frac{1}{3}r^2}$ orbitals for the $\beta$ = 1 to 9, respectively. The contribution in the tunnelling matrix element from each tip orbital is based on the derivatives defined in table~\ref{tab:table1}. The weights of the orbital contributions in the overall tip state are then optimized to quantitatively match the experimental measurement. In all of our STM measurements (more than twenty measured images of P and As are examined), we find dominant ($>$ 90\%) component from the $d_{z^2 - \frac{1}{3}r^2}$ orbital in the tip state. We infer it to the presence of high frequency components in the Si-dopant system, primarily arising from the valley quantum interference processes~\cite{Salfi_NatMat_2014}. Interestingly, while the $s$ orbital tip can capture low frequency components~\cite{Salfi_NatMat_2014}, a high level agreement and hence precision metrology can only be achieved if higher order tip orbitals are included. 

\begin{table}
\caption{\label{tab:table1} Relationship of the tunnelling matrix element $M_ \textrm {DT}$ and the dopant wave function $\Psi_ \textrm D$ is defined by a functional whose definition depends on the type of orbital in the STM tip in accordance with the derivative rule proposed by Chen.~\cite{Chen_PRB_1990} Here r$_0$ is the position of STM tip above the sample surface.}
\begin{ruledtabular}
\begin{tabular}{cc}
Tip Orbital Type &  $\Im [\Psi_ \textrm D (r)]$, computed at r$_0$ \\
\hline
$s$ & $\Psi_ \textrm D(r)$ \\ \\
$p_x$ & $\frac{\partial \Psi_ \textrm D (r)}{\partial x}$ \\ \\
$p_y$ & $\frac{\partial \Psi_ \textrm D (r)}{\partial y}$ \\ \\
$p_z$ & $\frac{\partial \Psi_ \textrm D (r)}{\partial z}$ \\ \\
$d_{xy}$ & $\frac{\partial^2 \Psi_ \textrm D (r)}{\partial x \partial y}$ \\ \\
$d_{zy}$ & $\frac{\partial^2 \Psi_ \textrm D (r)}{\partial z \partial y}$ \\ \\
$d_{zx}$ & $\frac{\partial^2 \Psi_ \textrm D (r)}{\partial z \partial x}$ \\ \\
$d_{x^2 - y^2}$ & $\frac{\partial^2 \Psi_ \textrm D (r)}{\partial x^2} - \frac{\partial^2 \Psi_ \textrm D (r)}{\partial y^2}$ \\ \\
$d_{z^2 - \frac{1}{3}r^2}$ & $\frac{2}{3}\frac{\partial^2 \Psi_ \textrm D (r)}{\partial z^2} - \frac{1}{3}\frac{\partial^2 \Psi_ \textrm D (r)}{\partial y^2} - \frac{1}{3}\frac{\partial^2 \Psi_ \textrm D (r)}{\partial x^2}$ \\

\end{tabular}
\end{ruledtabular}
\end{table}

For the $d_{z^2 - \frac{1}{3}r^2}$ orbital, we calculate the derivatives of the dopant wave function at the tip location, where the vacuum decay of the dopant wave function is based on the Slater orbital real-space dependence~\cite{Slater_PR_1930}, which satisfies the vacuum Schr\"{o}dinger equation. Each calculated dopant image is comprised of 8$\times$8 nm$^2$ grid, consisting of $\sim$1886 pixels. The amplitude of each pixel is calculated by the summation of wave function contributions from the atoms within the $(3a_0)^3$ area. Each atom is represented by twenty orbitals in our tight-binding basis, so in total each pixel of the image is constructed by adding 4,320 orbital contributions. 

In order to carry out a quantitative analysis, the theoretically computed normalised images of the dopant wave functions are compared directly to the experimental images (similarly normalised) by computing the two comparator parameters described as follows: A direct comparison of images is performed by defining pixel-by-pixel difference comparator $C_{\rm P}$, where the difference of the experimental image pixels ($I_{\rm exp}(x,y)$) and the theoretical image pixels ($I_{\rm th}(x,y)$) is computed across the identical scanned areas ($A_{\rm scan}$=$N_x \times N_y$, where $N_x$ and $N_y$ are the total numbers of pixels along the $x$ and $y$ directions) and normalised by the total intensity of the experimental image, $i.e.$:

\begin{equation}
C_{\rm P} = 1 - \left( \frac{\sum \limits_{x,y}^{N_x,N_y} |I_{\rm exp}(x,y) - I_{\rm th}(x,y) |}{\sum \limits_{x,y}^{N_x,N_y} I_{\rm exp}(x,y)} \right)
\end{equation} 

Although $C_{\rm P}$ is a direct comparison of the pixels in two images, it does not take into account the structure of images, $i.e.$ number, size, and intensity of the features inside an image. Therefore, we also define feature-by-feature correlation comparator $C_{\rm F}$, which takes into account both the structural information of the images and the intensity of the individual features in the images. The definition of the $C_{\rm F}$ comparator is as follows:

\begin{equation}
C_{\rm F} = \left( 1 - \frac{|N_{\rm F}^{exp} - N_{\rm F}^{th}|}{N_{\rm F}^{exp}} \right) \left( \frac{\sum\limits_{x,y}^{N_{\rm F}^{exp}} \sum\limits_{x,y}^{N_{\rm F}^{th}} [I_{\rm th}(x,y) I_{\rm exp}(x,y)]}{\sqrt{\sum\limits_{x,y}^{N_{\rm F}^{th}}I_{\rm th}^2(x,y)\sum\limits_{x,y}^{N_{\rm F}^{exp}}I_{\rm exp}^2(x,y)}} \right)
\end{equation} 
\noindent
where $N_{\rm F}^{th}$ and $N_{\rm F}^{exp}$ are total number of features in the computed and measured images, respectively, and $I_{\rm th}(x,y)$ and $I_{\rm exp}(x,y)$ denote the intensities of the $(x,y)$th pixels in the computed and measured image features, respectively. Note that the first term in $C_{\rm F}$ compares the structure of the two images, which would be 1.0 only if the two images have same number of features. The second term is a direct correlation of the pixel intensities inside the features and therefore would be 1.0 only if each corresponding feature has the same intensity. Overall a larger value of the $C_{\rm F}$ comparator -- the product of the two terms -- is a direct measure of the similarity between a theory image and an experimental image. In our quantitative analysis of all of the four measured data sets (Fig. 3 and Fig.~\ref{fig:EFigure8}), the allocation of donor positions is based on the concurrence of peaks for both $C_{\rm P}$ and $C_{\rm F}$ parameters. 

\begin{center}
\textcolor{blue}{
\textbf{S4. Two-step procedure to assign a location group to a measured STM image} } \\
\end{center}

We have developed a systematic recipe to find the exact position of dopant atom corresponding to a measured STM image. The procedure is based on the symmetry analysis of the bright features in the STM images in two steps. Fig.~\ref{fig:EFigure4} shows the calculated P images as a function of $n$ with each column corresponding to one of the six location-groups. Overall with respect to the [-110] axis, we find that the images for the dopants placed in the $L_{3/4}^i(n)$ and $L_{0}^i(n)$ groups exhibit one central bright lobe, independent of lateral positioning $i$. On the other hand, the dopants placed in the $L_{1/4}^i(n)$ and $L_{1/2}^i(n)$ groups demonstrate two side lobes in the images (butterfly symmetry), again irrespective of the lateral positioning. This is further highlighted in Fig.~\ref{fig:EFigure5} for $n$=7 by using the dashed lines and ovals. In accordance with an initial classification in Ref.~\onlinecite{Salfi_NatMat_2014} the dopant images with one(two) central(side) lobe(s) are labelled as $B$($A$)-type images (Fig.~\ref{fig:EFigure4}). Therefore our theory reveals that the dopants placed in the $L_{1/4}^i(n)$ and $L_{1/2}^i(n)$ location-groups produce $A$-type images, whereas $B$-type images correspond to the dopants located in the $L_{3/4}^i(n)$ and $L_{0}^i(n)$ location-groups, following an overall pattern of $A, A, B, B, A, A,$... as $d[P_{m}(n)]$ increases. In the first step -- without any theoretical calculation -- by just associating a type to a measured image, half of the six possible location-groups could be eliminated.
 
In the second step, the symmetry of the bright features across the [110]-diagonal plotted through the centre of the image is examined. This uniquely associates a single location-group to a dopant image as described in Fig.~\ref{fig:EFigure6}. In conclusion -- merely based on the symmetry properties -- the above-mentioned two-step recipe uniquely allocates one location-group to a measured image. 

To demonstrate the working of the procedure, we have applied it to the four experimentally measured images (P-1-Exp, P-2-Exp, As-1-Exp, As-2-Exp) as shown by a flow chart diagram illustrated in Fig.~\ref{fig:EFigure7}. The two-step procedure uniquely allocates $L_{3/4}^7(n)$, $L_{3/4}^8(n)$, $L_{0}^{1,2}(n)$, and $L_{1/2}^6(n)$ locations groups to the measured images P-1-Exp, P-2-Exp, As-1-Exp, and As-2-Exp, respectively.

\begin{center}
\textcolor{blue}{
\textbf{S5. Determining STM image type ($A$ or $B$) from the Fourier sprectum}} \\
\end{center}

In Fig.~\ref{fig:EFigure5}, we have shown that a dopant image can be assigned a type ($A$ or $B$) based on the symmetry of the bright features. Here we show that the type ($A$ or $B$) of the dopant images is also consistently reflected in the FTS features along the $k_x$=$k_y$ line (marked by an arrow in Fig.~\ref{fig:EFigure2}. Type-$A$ dopants exhibit a sharp decrease in the Fourier amplitude between the $k$=0 peak and the $k_x$=$k_y$=0.15(2$\pi$/$a_0$) peak, which is absent for the type-$B$ dopants~\cite{Salfi_NatMat_2014}. Both the measured and calculated FTS shown in Fig.~\ref{fig:EFigure2} do not exhibit a sharp decrease in the FT amplitude along the arrow direction, consistent with the $B$-type determined from one central lobe of the bright features in the real-space image. 

Interestingly a direct FTS comparison of images calculated with $s$ and $d_{z^2- \frac{1}{3}r^2}$ tips in Fig.~\ref{fig:EFigure9}  reveals a flip of the image types -- $s$-wave tip would measure an $A$-type dopant (from the $d_{z^2- \frac{1}{3}r^2}$ tip) as a $B$-type dopant and vice versa. Therefore we conclude that proper resolution of tip state is a crucial requirement to understand Fourier spectrum of the measured STM images, in particular the high frequency components.

\renewcommand{\thefigure}{S\arabic{figure}}
\setcounter{figure}{0}

\begin{figure*}
\centering
\includegraphics[scale=0.15]{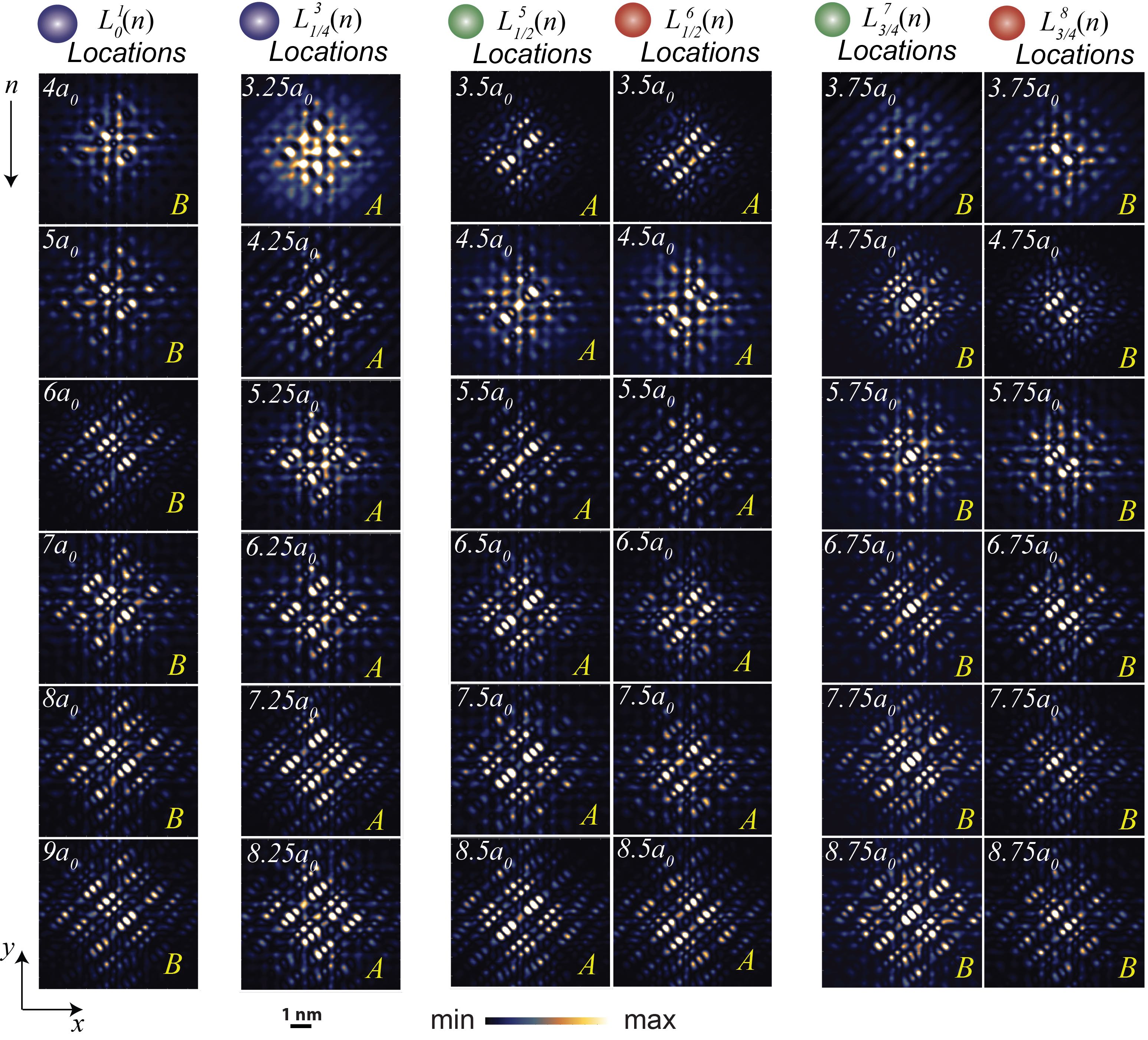}
\caption{\textbf{Calculated images categorised in location groups as a function of $n$:} Calculated images of a P dopant are plotted by using a tip height of 2.5$\textrm{\AA}$. Each column represents one of the six possible location groups for dopant placement, and the images within the column are obtained by placing dopants at individual positions (by increasing $n$). The depth of the location and the type of the dopant ($A$ or $B$) at that location is also marked in accordance with the procedure defined in Fig.~\ref{fig:EFigure5}.}
\label{fig:EFigure4}
\end{figure*}

\newpage
\begin{figure*}
\centering
\includegraphics[scale=0.15]{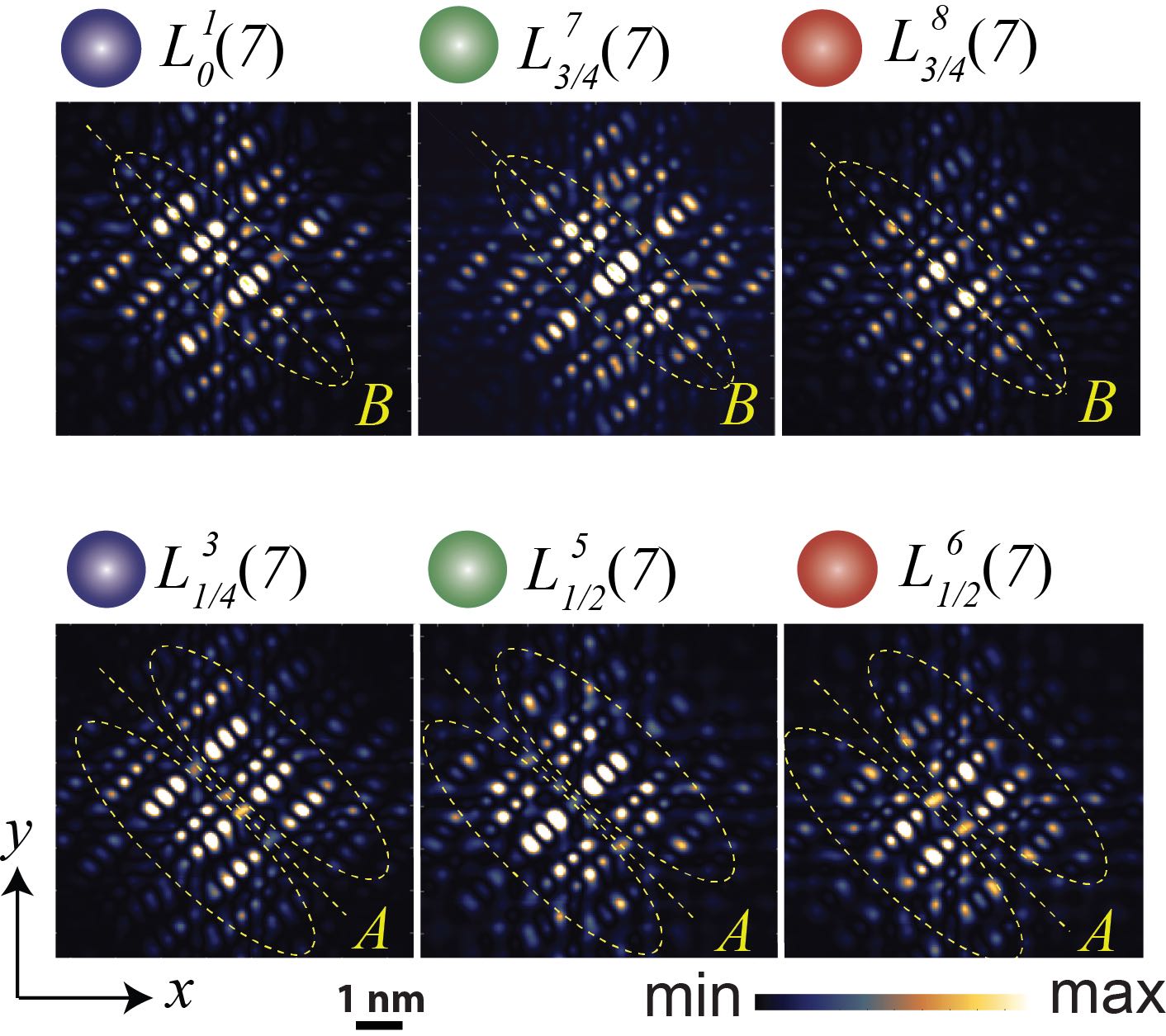}
\caption{\textbf{Determining the type ($A$ or $B$) of a dopant image:} As an example, we select $n$=7 case from the Fig.~\ref{fig:EFigure4}. With respect to the [-110] diagonal (dashed line), the presence of one central or two side lobes of the bright features is highlighted by plotting the dashed yellow ovals. Each image is assigned a type based on the count of lobes: one central lobe = type $B$ image, two side lobes = type $A$ image.}
\label{fig:EFigure5}
\end{figure*}

\newpage
\begin{figure*}
\centering
\includegraphics[scale=0.2]{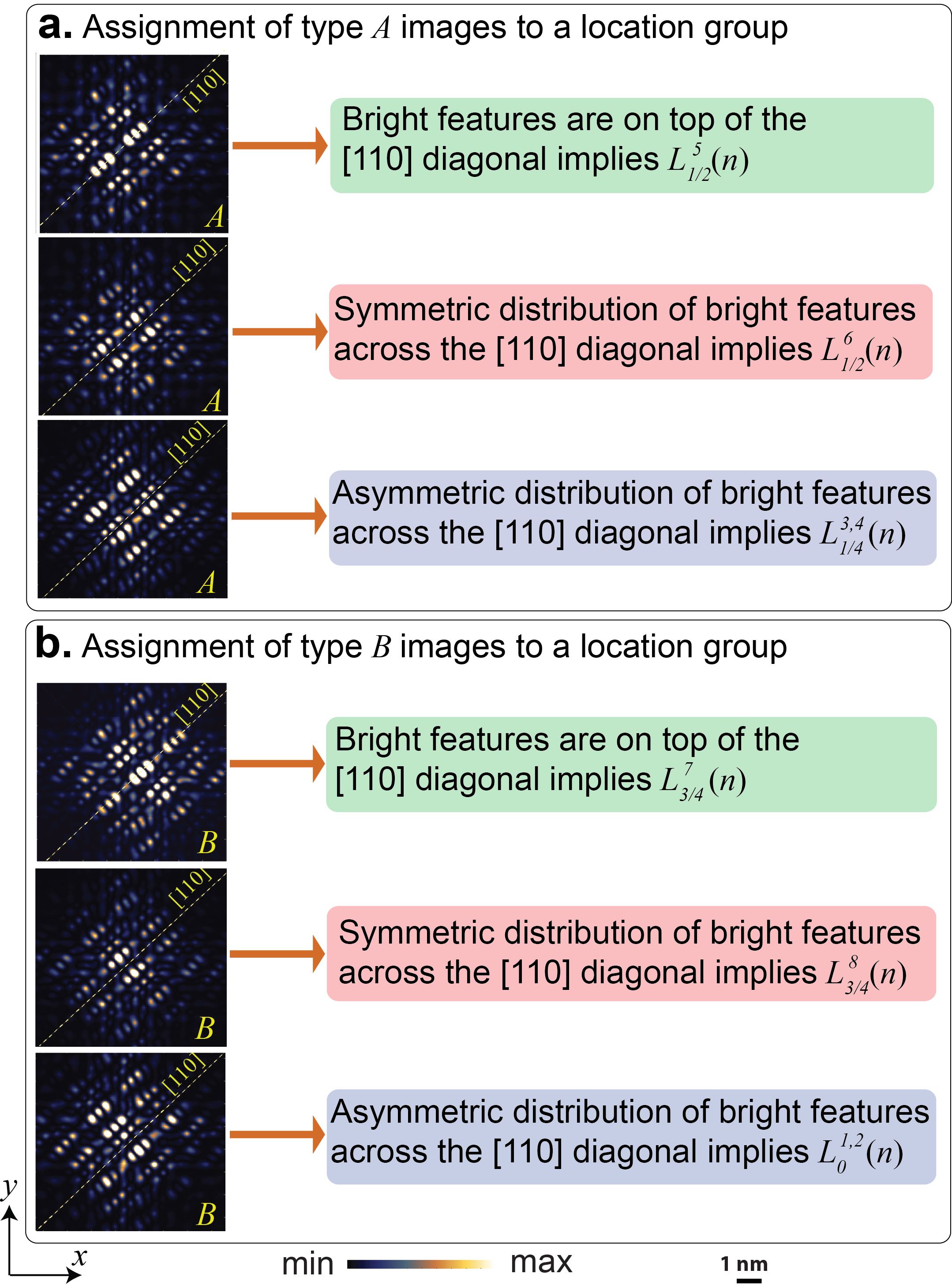}
\caption{\textbf{Associating a location-group to $A$/$B$ type of the dopant image:} Methodology to assign a location-group (finding the values of $m$ and $i$ in $L_{m}^i(n)$) is presented, when the type of the image is already known from the procedure described in Fig.~\ref{fig:EFigure5}. The symmetry of the bright features is examined with respect to the [110] diagonal plotted through the center of the image, which leads to a unique assignment of one of the six location-groups. The boxes on the right side are color coded in accordance with the color coding of the corresponding location-groups defined in Fig. S8.}
\label{fig:EFigure6}
\end{figure*}

\newpage
\begin{figure*}
\centering
\includegraphics[scale=0.1]{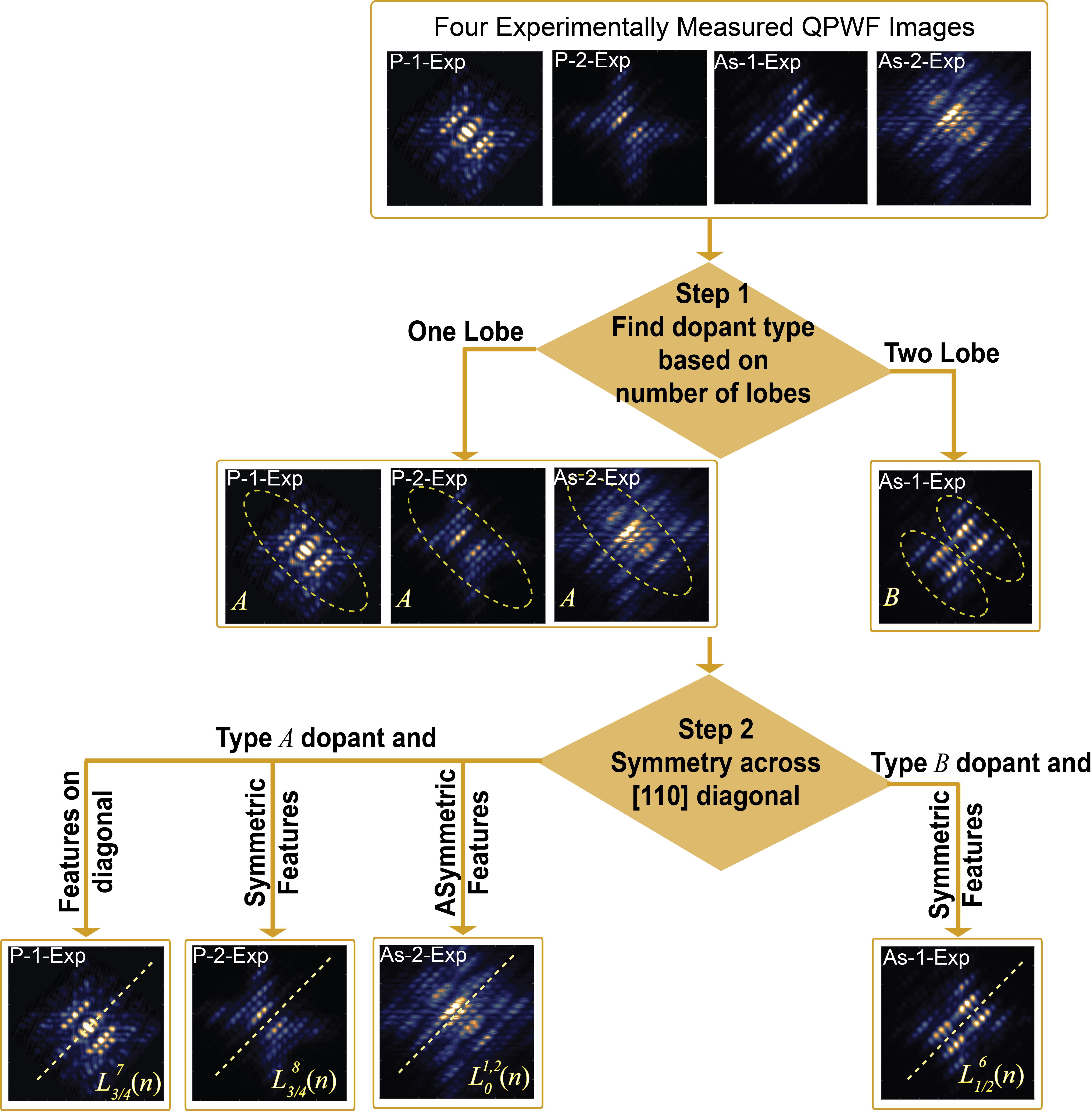}
\caption{\textbf{Flow chart of the two-step procedure applied to the STM measured images:} The two-step procedure to find a location group $L_{m}^i(n)$ for the experimentally measured images (P-1-Exp, P-2-Exp, As-1-Exp, and As-2-Exp) is demonstrated with the help of flow-chart diagram. In the first step, type ($A$ or $B$) is assigned to the dopant images based on the number of bright feature lobes with respect to [-110] diagonal. The second step examines symmetry of the bright features across the [110] diagonal plotted through the center of the image, and allocates a unique location group to the image based on the recipe outlined in Fig.~\ref{fig:EFigure6}.}
\label{fig:EFigure7}
\end{figure*}

\newpage
\begin{figure*}
\centering
\includegraphics[scale=0.15]{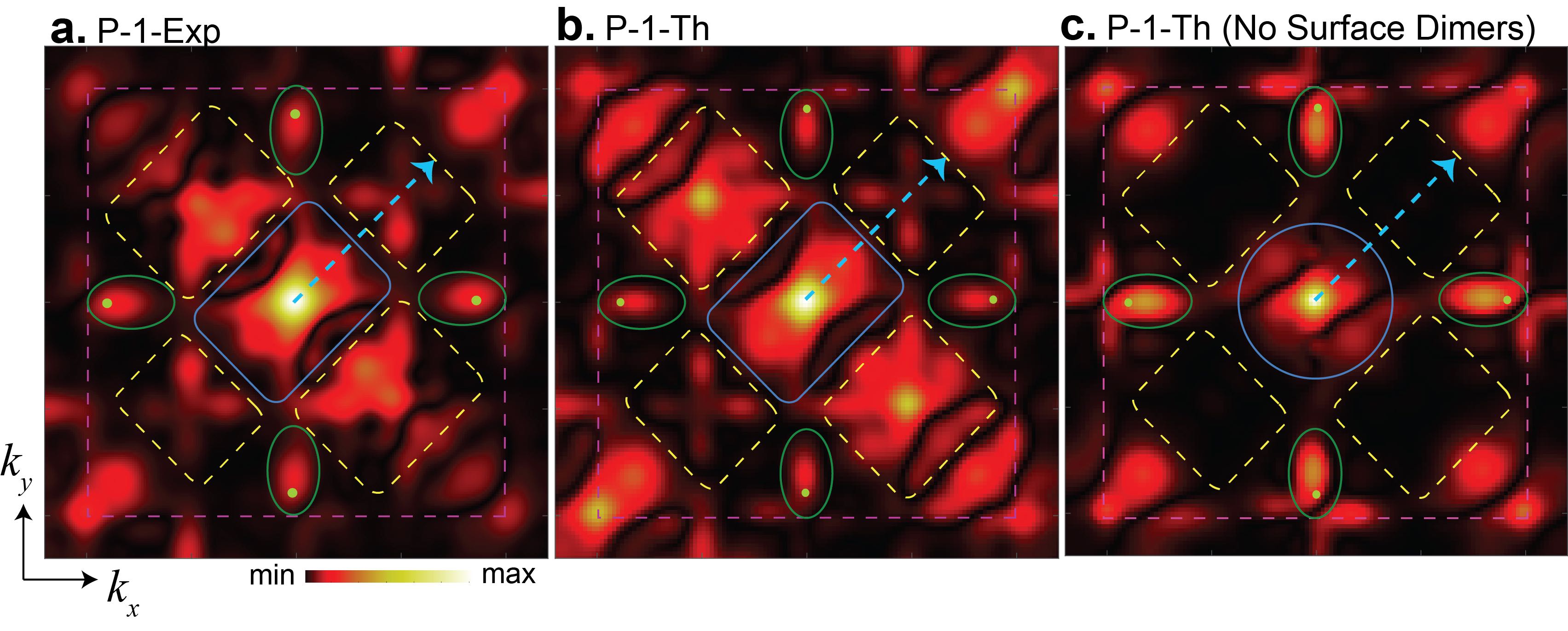}
\caption{\textbf{Fourier transform spectra of measured and calculated P-1 image:} \textbf{a.} Fourier transform spectrum of the experimentally measured P-1-Exp dopant image. The corners of the outer dashed purple box are reciprocal lattice vectors (2$\pi$/$a_0) (p,q)$, with $p$=$\pm$1 and $q$=$\pm$1. The ellipsoidal structures corresponding to valleys are found within the green ovals and the green dots indicate the position of the conduction band minima: $k_x$ = 0.85(2$\pi$/$a_0) (\pm1,0)$ and $k_y$ = 0.85(2$\pi$/$a_0) (0,\pm1)$. The region marked with blue boundary indicates probability envelope and the yellow dashed regions highlight the reconstruction-induced features. \textbf{b.} The Fourier transform of the calculated dopant image P-1-Th, with the dopant depth 4.75$a_0$, tip height as 2.5$\textrm{\AA}$, $d_{z^2- \frac{1}{3}r^2}$-wave tip. \textbf{c.} The calculated image with the same configuration as (b) but the 2$\times$1 surface reconstruction is switched off. Note the absence of reconstruction related features within the yellow dashed regions and the central region around $k_x$=$k_y$=0 is also symmetric.}
\label{fig:EFigure2}
\end{figure*}

\newpage
\begin{figure*}
\centering
\includegraphics[scale=0.15]{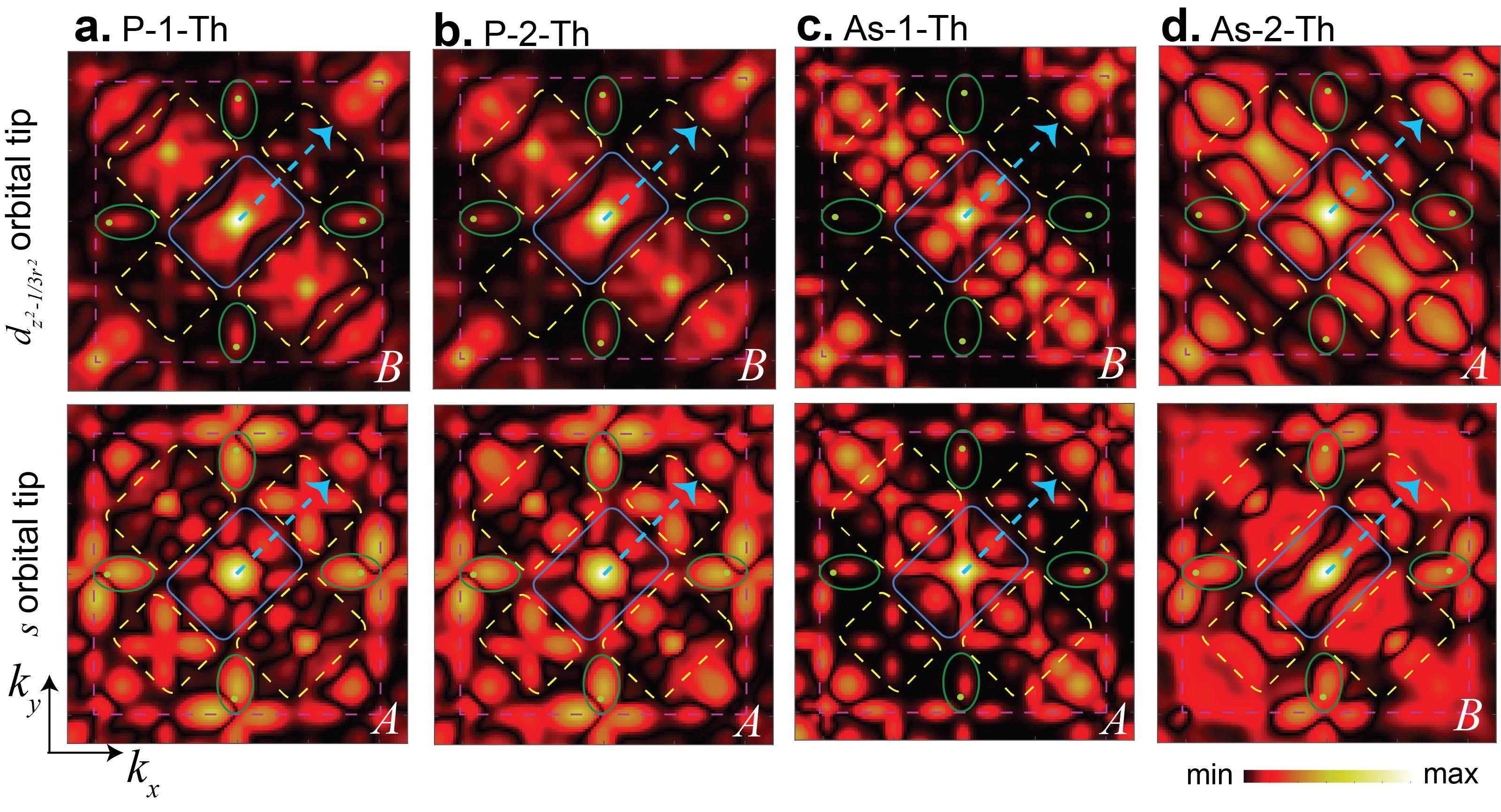}
\caption{\textbf{Effect of tip orbitals on the Fourier transform spectra of calculated images:} Fourier transform spectrum of the theoretically calculated images: \textbf{a.} P-1-Th, \textbf{b.} P-2-Th, \textbf{c.} As-1-Th, and \textbf{d.} As-2-Th. The corners of the outer dashed purple box are reciprocal lattice vectors (2$\pi$/$a_0) (p,q)$, with $p$=$\pm$1 and $q$=$\pm$1. The ellipsoidal structures corresponding to valleys are found within the green ovals and the green dots indicate the position of the conduction band minima: $k_x$ = 0.85(2$\pi$/$a_0) (\pm1,0)$ and $k_y$ = 0.85(2$\pi$/$a_0) (0,\pm1)$. The region marked with blue boundaries indicates probability envelope. The reconstruction-induced features are highlighted with the yellow dashed regions. The two rows compare $d_{z^2- \frac{1}{3}r^2}$ orbital (upper row) and $s$ orbital (lower row) tip configurations.}
\label{fig:EFigure9}
\end{figure*}

\begin{figure*}
\centering
\includegraphics[scale=0.1]{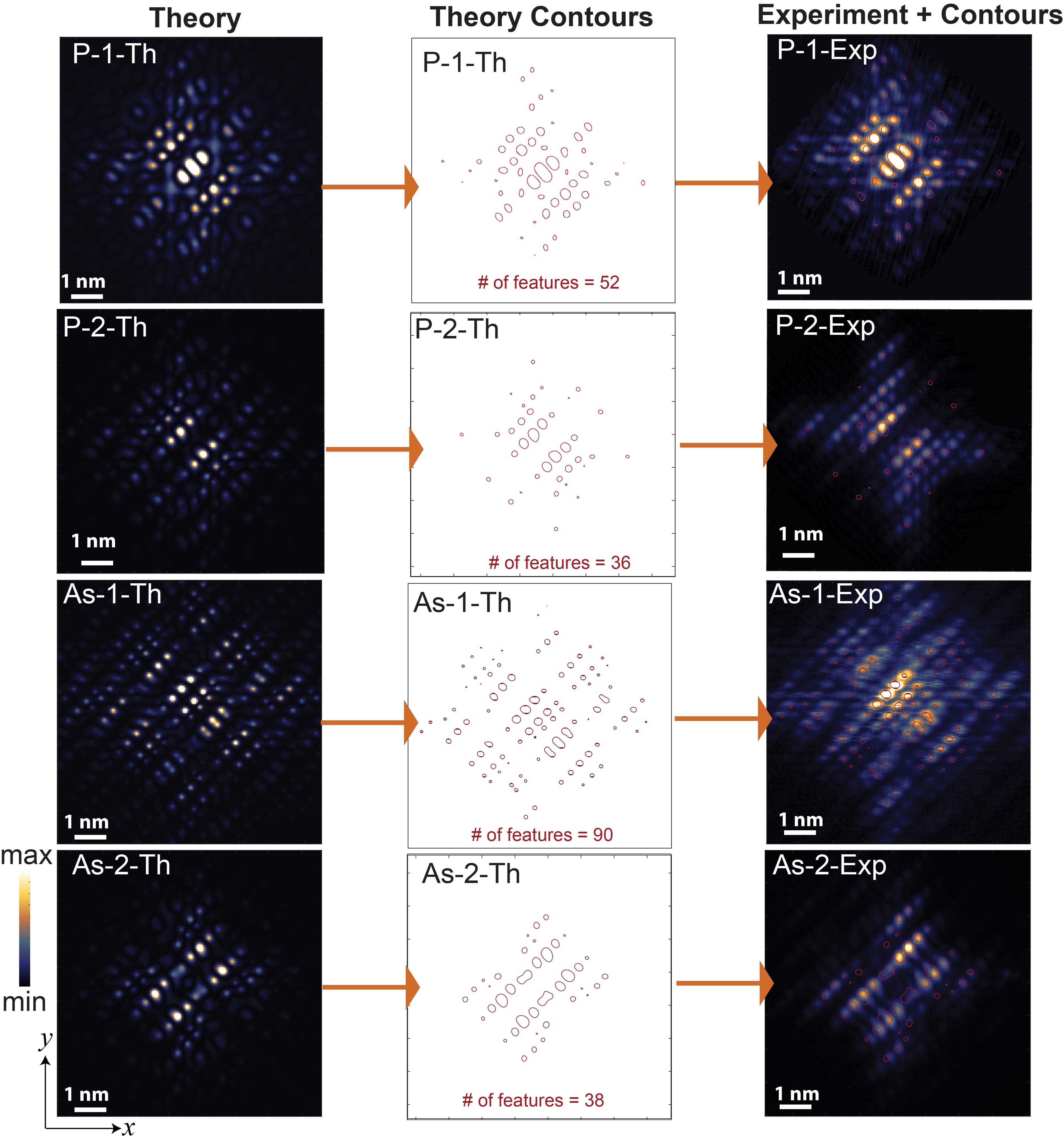}
\caption{\textbf{Feature-by-feature correlation between measured and calculated P and As images:} (Left Column) Theoretically computed images of the four dopants P-1, P-2, As-1, and As-2 using $d_{z^2- \frac{1}{3}r^2}$ orbital tip. (Middle Column) Contours of the bright features extracted from the calculated images. (Right Column) Experimentally measured dopant images with contours of the calculated images overlaid to highlight one-to-one correspondence between the features of experimental and theoretical images.}
\label{fig:EFigure1}
\end{figure*}

\newpage
\begin{figure*}
\centering
\includegraphics[scale=0.15]{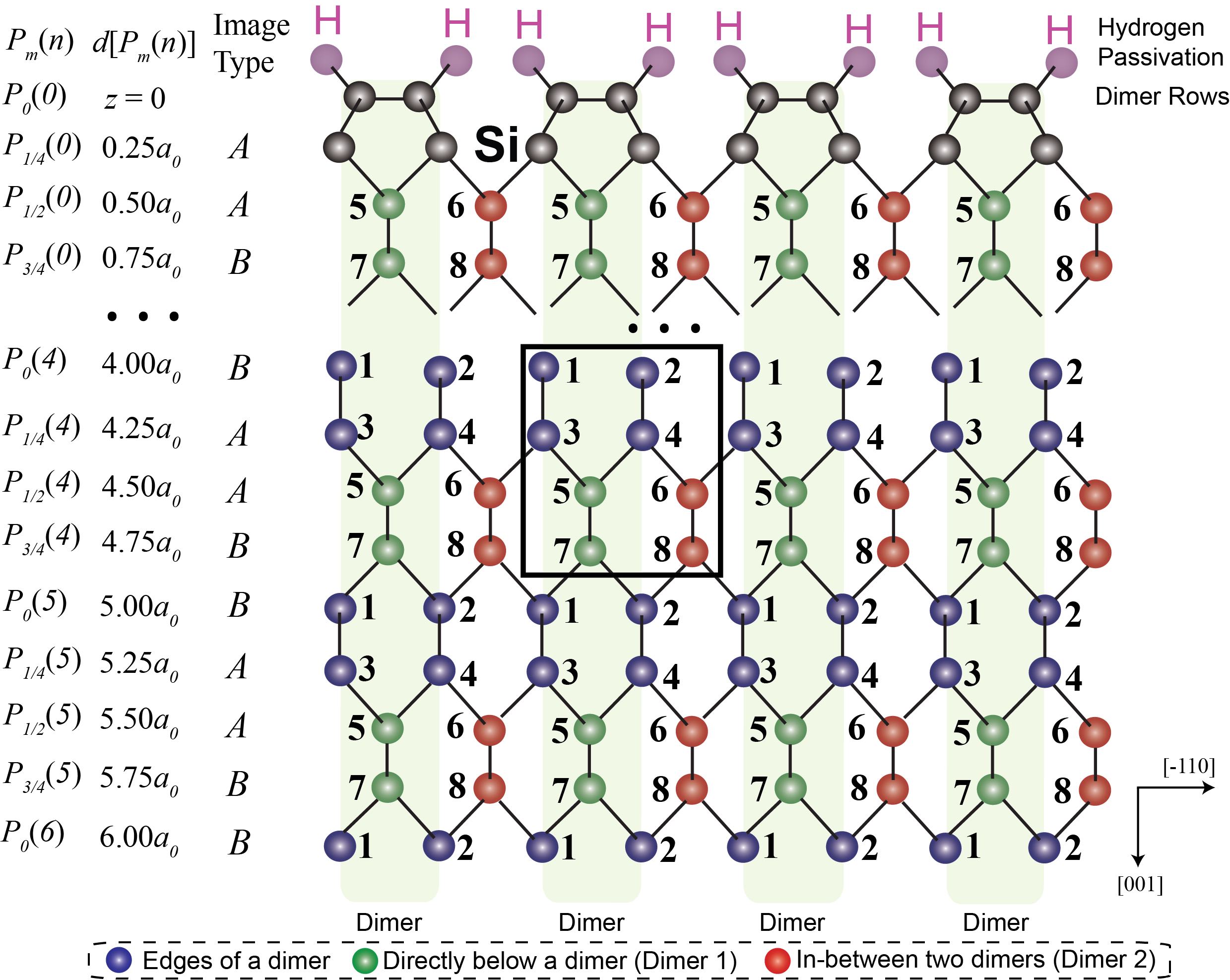}
\caption{\textbf{Nomenclature of possible atomic sites for dopant:} The schematic diagram of a small portion of the Si crystal structure is shown to illustrate the possible locations for a dopant atom within a few nanometers of the $z$=0 surface. The $z$=0 surface is Hydrogen passivated (purple atoms) and exhibits the formation of Si dimer rows (light blue atoms), which are aligned perpendicular to the page (along the [110] direction). The area shaded underneath the dimers is to guide the eyes on the positioning of atomic sites with respect to the dimers. A systematic labelling scheme is designed by coupling the symmetries of the dopant images with the periodicity of the Si lattice. Along the [001] direction, Si lattice planes are divided into four plane groups ($PG_m$, where $m \in$ \{0, 1/4, 1/2, 3/4\}). Each plane group is a set of planes $P_{m}(n)$, whose depths from the $z$=0 surface are given by: $d[P_{m}(n)]$=($m$+$n$)$a_0$, where $n$=0,1,2,3,... Note that $(m,n)$=(0,0) represents the $z$=0 surface. Considering the periodicity of the dimers along the [-110] direction, two possible dopant locations $L_{m}^i(n)$ per plane $P_{m}(n)$ are defined by $i=8m+1$ and $i=8m+2$, which repeat periodically in the lateral direction. The coloring of the atomic sites is performed with respect to their positioning under the dimer rows along the [-110] axis. The positions within the $L_{1/2}^i(n)$ and $L_{3/4}^i(n)$ groups are either directly underneath the dimers marked with green color or in-between the two dimers marked with red color. As the lateral positions within the $L_{0}^i(n)$ and $L_{1/4}^i(n)$ groups are always at the edges of the dimers, so these are equivalent with respect to the surface strain and therefore are marked by a single blue color. Finally, we highlight a supercell (marked by solid blue boundary) with atomic positions labelled with the corresponding $i$ values. This supercell defines the overall periodicity of the dopant positions in the three dimensional lattice of Si.}
\label{fig:EFigure3}
\end{figure*}

\newpage
\begin{figure*}
\centering
\includegraphics[scale=0.2]{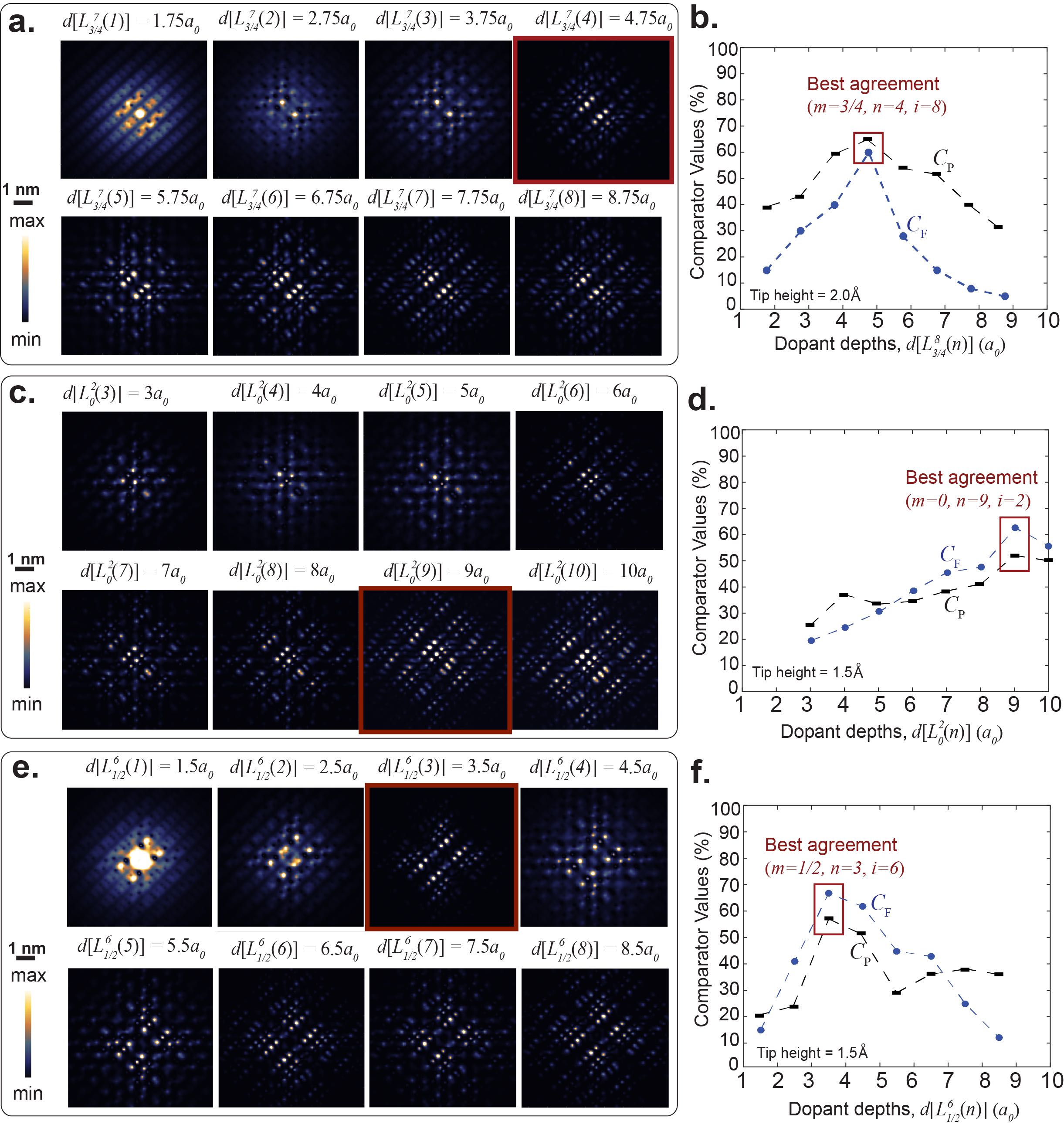}
\caption{\textbf{Identification of the unique lattice sites for P-2, As-1, and As-2 dopants:} \textbf{a.} Quantitative comparison between the experimentally measured and theoretically computed images is shown for P-2 dopant in (\textbf{a,b}), As-1 dopant in (\textbf{c,d}), and As-2 dopant in (\textbf{e,f}) First a location group is finalised for each experimental image from the procedure shown in Fig. S4. Here in each panel, we plot all the computed images within the location group on left side, and highlight one image, which match the experimental measurement. On the right side, we quantitatively compare measured and computed images by plotting comparator $C$ graphs, and uniquely identify a dopant location from the peak of the graph. It is noted that the case shown in \textbf{d.} highlights the limit of our procedure which is accurate up to around 36 atomic planes from the $z$=0 surface. For the deeper dopant locations, the images start converging towards bulk case and therefore making it hard to clearly distinguish the individual features.}
\label{fig:EFigure8}
\end{figure*} 

\clearpage
\newpage


%

\end{document}